\documentclass[journal]{IEEEtran}
\ifCLASSOPTIONcompsoc
  \usepackage[nocompress]{cite}
\else
  \usepackage{cite}
\fi
\ifCLASSINFOpdf
   \usepackage[pdftex]{graphicx}
\else
\fi

\usepackage{amsmath,amssymb,amsfonts}
\usepackage{mathtools, mathrsfs}
\usepackage{algorithm}
\usepackage[noend]{algpseudocode}
\usepackage{graphicx}
\usepackage{textcomp}
\usepackage{xcolor}
\usepackage{array}
\usepackage[flushleft]{threeparttable}
\usepackage{microtype}
\usepackage{url}
\usepackage{hyperref}
\usepackage[section]{placeins}

\DeclareMathOperator*{\argmin}{arg\,min}

\newcommand{\etal}{\emph{et al. }}
\mathchardef\mhyphen="2D

\begin{document}

\title{Efficient Cyber Attack Detection in Industrial Control Systems Using Lightweight Neural Networks and PCA}

\author{Moshe~Kravchik
        and~Asaf~Shabtai
\thanks{The authors are with the Department
of Software and Information Systems Engineering, Ben-Gurion University of the Negev, Beer-Sheva, Israel. E-mail: moshekr@post.bgu.ac.il, shabtaia@bgu.ac.il.}
}

\maketitle

\begin{abstract}
Industrial control systems (ICSs) are widely used and vital to industry and society. 
Their failure can have severe impact on both economics and human life. 
Hence, these systems have become an attractive target for attacks, both physical and cyber. 
A number of attack detection methods have been proposed, however they are characterized by a low detection rate, a substantial false positive rate, or are system specific.
In this paper, we study an attack detection method based on simple and lightweight neural networks, namely, 1D convolutions and autoencoders.
We apply these networks to both the time and frequency domains of the collected data and discuss pros and cons of each approach.
We evaluate the suggested method on three popular public datasets and achieve detection rates matching or exceeding previously published detection results, while featuring small footprint, short training and detection times, and generality.
We also demonstrate the effectiveness of PCA, which, given proper data preprocessing and feature selection, can provide high attack detection scores in many settings.
Finally, we study the proposed method's robustness against adversarial attacks, that exploit inherent blind spots of neural networks to evade detection while achieving their intended physical effect.
Our results show that the proposed method is robust to such evasion attacks: 
in order to evade detection, the attacker is forced to sacrifice the desired physical impact on the system.
This finding suggests that neural networks trained under the constraints of the laws of physics can be trusted more than networks trained under more flexible conditions.
\end{abstract}



\section{Introduction}\label{sec:introduction}
Industrial control systems (ICSs), also known as supervisory control and data acquisition (SCADA) systems, combine distributed computing with physical process monitoring and control. 
They are comprised of elements providing feedback from the physical world (sensors), elements influencing it (actuators), as well as computers and controller networks that process the feedback data and issue commands to the actuators.
Many ICSs are safety-critical, and an attack interfering with their functionality can cause substantial financial and environmental harm, and endanger people's lives.

The importance of ICSs makes them an attractive target for attacks, particularly cyber attacks. 
Several high impact incidents of this kind have been reported in recent years, including the attack on a power grid in Ukraine~\cite{web:Ukrain_power_hack}, the infamous Stuxnet malware that targeted nuclear centrifuges in Iran~\cite{kushner2013real}, and attacks on a Saudi oil company~\cite{web:Saudi_Aramco_hack}.
In the past, ICSs ran on proprietary hardware and software in physically secure locations, but more recently they have adopted common information technology (IT) stack and remote connectivity. 
This trend exposes ICSs to cyber threats that leverage common technology stack attack tools.
At the same time, the ICS defender's toolbox is limited due to the need to support legacy protocols built without modern security features, as well as the inadequate processing capabilities of the endpoints.
This problem can be addressed by utilizing traditional IT network-based intrusion detection systems (IDS) to identify malicious activity, which does not rely on endpoint computational resources. 
However, the very low number of known attacks on ICSs renders this approach ineffective.
Alternatively, model-based methods have been proposed to detect anomalous behavior of the monitored ICS~\cite{pasqualetti2011cyber, teixeira2012attack, jones2014anomaly, mishra2019modeling}.
Unfortunately, creating an accurate model of complex physical processes is a very challenging task.
It requires an in depth understanding of the system and its implementation, and is time consuming and difficult to scale. 
Thus, recent studies have utilized machine learning to model the system. 
Some of them used supervised learning~\cite{beaver2013evaluation, hink2014machine} and achieved high precision results, however the supervised learning approach is limited to the modeled attacks only.
To overcome this obstacle, a number of other studies used unsupervised deep neural networks (DNNs) for detecting anomalies and attacks in ICS data~\cite{goh2017anomaly, inoue2017anomaly, lin2018tabor}.
Kravchik and Shabtai~\cite{kravchik2018detecting} suggested using unsupervised neural networks based on 1D CNNs and demonstrated the detection of 31 out of 36 cyber attacks in the popular SWaT dataset~\cite{goh2016dataset}, improving upon previously published results.
This paper extends the research performed in  ~\cite{kravchik2018detecting} and is aimed at answering the following questions that were not addressed in that study.
\begin{itemize}
    \item Can the generality and effectiveness of a 1D CNN be validated using additional datasets, preferably from different types of system?
    \item Are there alternative lightweight neural network architectures that can be used in the method proposed in ~\cite{kravchik2018detecting}?
    \item Can the detection of an anomaly be interpreted in such a way that it provides actionable insights to the system operator, namely how to pinpoint the attacked sensors or actuators?
    \item Will detection in the frequency domain provide any benefits compared to detection in the time domain?
    \item How robust are the proposed neural networks architectures to adversarial machine learning attacks?
\end{itemize}

The contributions of this paper are as follows.
\begin{itemize}
    \item An effective and generic method for detecting anomalies and cyber attacks in ICS data using 1D CNNs and undercomplete autoencoders (UAEs).
    The method was validated on three public datasets and achieved better performance than previously published research in this area.
    \item A method for robust feature selection based on the Kolmogorov-Smirnov test.
    \item A method for attack detection \textbf{explanation} - attack localization using neural network-based models.
    \item The efficient application of the above mentioned detection method to the frequency domain which provides high detection scores and guidelines for when to use it.
    \item A demonstration of the adversarial robustness of the proposed method under a powerful white-box attacker threat model.
\end{itemize}

The rest of this paper is organized as follows. 
In Section~\ref{sec:background}, we present the necessary background on ICSs, the considered threat model, relevant neural network architectures, time-frequency transformation, and adversarial machine learning.
In Section~\ref{sec:related}, we survey related work, focusing on physical system state-based detection research. 
Section~\ref{sec:datasets}, introduces the datasets we used for validation. 
We describe our methodology in Section~\ref{sec:method}. 
The experiments and their results are described in Section~\ref{sec:experiments}, and finally, Section~\ref{sec:conclusions} concludes the paper.

\section{Background}\label{sec:background}
\subsection{Industrial Control Systems}\label{sec:background_ics}
A typical ICS combines network-connected computers with physical processes, which are both controlled by these computers and provide the computers with feedback.
The key components of an ICS include sensors and actuators that are connected to a local computing element, commonly called a programmable logic controller (PLC) or a remote terminal unit (RTU). 

Sensors and actuators are usually connected to the PLC with a direct cable connection, and commands are sent to the PLC via a local networking protocol, such as CAN, Profibus, DNP3, IEC 61850, IEC 62351, Modbus, or S7.
PLCs of the remote nodes are connected to the central control unit via protocols, such as TCP, over a wireless, cellular, or wired network. 
The connection is made with the help of a data acquisition system (DAS) that bridges the local and remote networking protocols.
The central control node contains a master terminal unit (MTU) that applies the control logic to the RTUs and provides management capabilities to a human machine interface (HMI) computer.
A historian server that collects and stores data from the RTUs is another important ICS component.
An additional common element of ICSs is an engineering workstation running SCADA software that provides a means to both monitor the PLCs and to change their internal logic.

Recently, many SCADA and PLC components started supporting the Common Industrial Protocol (CIP), which allows the integration of industrial control applications with a standard network stack.
Using SCADA and PLC components supporting such modern protocols enables system simplification, allowing the removal of dedicated MTU and DAS components.


\subsection{Attacks on ICSs and Threat Model}\label{sec:background_threat_model}
The central role of ICSs in critical infrastructure, medical devices, transport, and other areas of society makes them an attractive target for attacks. 
Motives for such attacks are diverse and include criminals seeking control of an important asset or blackmailing a victim, industrial espionage and sabotage, political reconnaissance, cyber war, and privacy evasion.

An ICS can be attacked using several attack vectors including software, hardware, communication protocols, the physical environment, and human elements. \begin{figure}[b!]
\centerline{\includegraphics[clip, trim=0cm 0cm 0cm 0cm, scale=0.6]{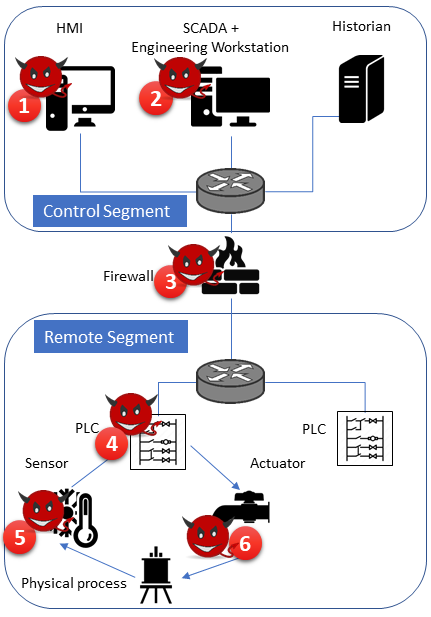}}
\caption{A schematic of ICS architecture and possible attack locations on an ICS system.}
\label{fig:ICS_attacks}
\end{figure}

For example, an adversary can attack:
\begin{enumerate}
    \item The HMI machine, by exploiting software vulnerabilities in its OS and application stack, presenting a fake view of the process and causing the operator to issue erroneous commands~\cite{kleinmann2017stealthy},
    \item The SCADA and/or engineering workstation machine, by exploiting software vulnerabilities and obtaining full control of the ICS, as it happened in Ukraine~\cite{web:Ukrain_power_hack},
    \item The communication network in the control segment, in the remote segment, or between them, by performing eavesdropping, or replay or false packet injection attacks,
    \item The PLC, by exploiting software vulnerabilities or trust between the PLC and SCADA; this will allow the attacker to change the PLC logic influencing the controlled process and cause damage, as in the Stuxnet case,
    \item The sensors, by leveraging physical effects interfering with the measurement or replacing the sensor with a malicious one as shown in~\cite{ahmed2018noise}, or,
    \item The actuators, by altering the signal sent by the actuators to the controlled process, as described in~\cite{giraldo2018survey}.
\end{enumerate}

The attack can also combine a number of vectors, for example issuing malicious commands to the actuator and replaying a valid system state to the SCADA, as done by the Stuxnet malware. 
Figure~\ref{fig:ICS_attacks} illustrates most common attack locations.

Our threat model considers a powerful adversary that is able to influence the physical state of the protected ICS. 
Regardless of the attack vector, the most common ultimate goal of the attacker is a physical-level process change.
Hence, in this research, we don't assume any specific attack vector, and apply a physics-based attack detection approach. 
The main idea of this approach is that the behavior of the protected system complies with immutable laws of physics and therefore can be modeled. 
Monitoring the physical system state and its deviation from the model facilitates the detection of anomalous behavior, including the deviations caused by spoofed sensor readings and injected control commands. 
For example, opening a valve should result in an increase in the water level. 
If the level does not increase or the speed of the increase is higher than usual, the sensor reading could have been falsified, or the sensor might be faulty.

Despite the fact that our anomaly detection domain is purely physical, we argue that our method goes \textbf{beyond simple anomaly detection}.
As we show, it can detect sophisticated multi-point cyber attacks that combine data tampering, malicious commands and replay attacks.
These attacks are targeted, evasive, and performed by means of the cyber domain.
Therefore, we choose to classify the method as \textbf{cyber attack detection}.
Adding the cyber context to our \textbf{detection domain}, e.g., by combining network and sensory data, is a promising direction for future research.
As the network data was not available in most of the public datasets, we made the decision to focus on physical-only detection in this research.

\subsection{Convolutional Neural Networks}\label{sec:background_cnn}
Convolutional neural networks (CNN) are feedforward neural networks popular in image processing domain.
In the basic neural network model, the layers are fully connected, which means that a unit (a neuron) is connected to all of the units in the subsequent layer. 
This requires the neuron to hold a very large number of weights on these connections.
This structure does not scale well, and such a large number of parameters (weights) will usually lead to overfitting. 
In addition, fully connected networks ignore the input topology: input variables can be presented in any order, and the outcome will be the same. 
However, many kinds of data, including images, have a distinct structure, and nearby pixels are highly correlated.
CNNs address these deficiencies by applying convolutions (filters) to small regions of the input instead of performing matrix multiplication on all of the input at once.  
The filter uses the same weights for all of the locations and thus can detect features regardless of their position in the image.
A convolutional layer consists of several feature maps each detecting a different input feature.
1D CNNs can successfully be used for time series processing, because time series have distinct 1D (time) locality that can be extracted by convolutions~\cite{lecun1995convolutional}.
\subsection{Autoencoders}\label{sec:autoencoders}
\begin{figure}[t!]
\centering{\includegraphics[clip, trim=0cm 0cm 0cm 0cm, scale=0.65]{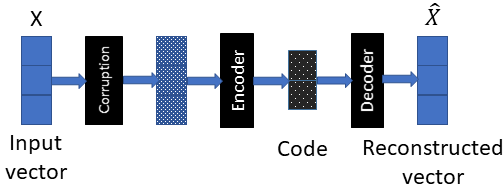}}
\caption{Denoising Autoencoder.}
\label{fig:DAE}
\end{figure}

An autoencoder (AE) is a neural network trained to reproduce its input~\cite{goodfellow2016deep}, thereby learning useful properties of the data.
This is achieved by applying constraints on the network which prevent copying the input to the output and cause the network to learn the compact representation of the data.
Due to this ability, autoencoders are widely used for dimensionality reduction and feature learning \cite{goodfellow2016deep}.
Autoencoders have two major components: an encoder that transforms the input into some internal representation, and a decoder that reconstructs the input from this representation.
The simplest kind of autoencoder is an \textit{undercomplete} autoencoder (UAE), which passes the data through a bottleneck of a hidden layer with smaller dimensions than the input and output.
This bottleneck forces the network to learn a subspace which captures principal features of the data. 
Another way of forcing an autoencoder to learn important input structural features is letting it reconstruct the original input from the input after it has been corrupted by noise.
A common way to corrupt the input is to add some Gaussian noise to it.
Autoencoders that utilize this technique are called \textit{denoising autoencoders} (DAEs) (see Figure~\ref{fig:DAE}).

\begin{figure}[t!]
\centering{\includegraphics[clip, trim=0cm 0cm 0cm 0cm, scale=0.375]{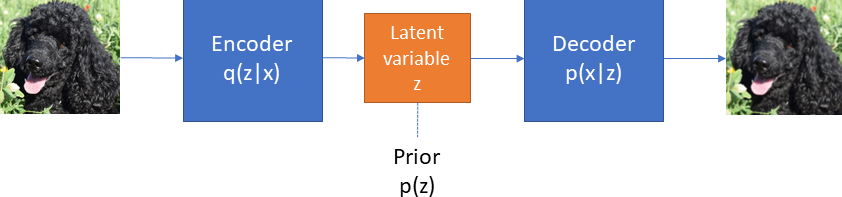}}
\caption{Variational Autoencoder.}
\label{fig:VAE}
\end{figure}

\textit{Variational} autoencoders (VAE) \cite{doersch2016tutorial} have become very popular in the unsupervised learning of complex distributions and in generating images of different kinds. 
While regular autoencoders learn a compact representation of the input data, there is no constraint on this compact representation.
For example, given an autoencoder network trained with many images of dogs, we still don't know how to build an internal representation that could generate a dog when passed to the decoder part of the network.
VAE solves this problem by applying constraints on the distribution of the compact representation (called a latent variable or code).
To impose this constraint, the loss function is a sum of the data reconstruction error (generalization error) and a deviation of the latent variable distribution from some chosen prior distribution, typically a unit Gaussian distribution.
Once the network is trained, it is possible to generate new images of dogs by drawing samples from the unit Gaussian distribution and passing them to the decoder.
A VAE has three parts: an encoder, decoder, and prior distribution (as illustrated in Figure~\ref{fig:VAE}).
The encoder creates a distribution of the latent variables for the given input, and the decoder returns a distribution of inputs corresponding to the given latent variables.
The network is trained to maximize the likelihood of the data given the codes it assigns to it, while maintaining the codes' distribution close to the chosen prior one.
\subsection{\label{subsec:freq_domain}Time - Frequency Domain Transformation}
Raw data measured by ICS sensors produces a time series.
While in most ICS anomaly and attack detection research this data is processed directly, it is very common in signal processing to analyze data in the frequency domain. 
Fourier transform \eqref{ft_continuous} allows us to build a signal's frequency domain representation:
\begin{equation}
  \label{ft_continuous}
  \hat{f}(k) = \int^{+\infty}_{-\infty} f(x)e^{-2 \pi i x k} dx,
\end{equation}
where $f$ is some function depending on time $x$, $\hat{f}$ is its Fourier transform, and $k$ is the frequency.
When dealing with periodic data samples, rather than a continuous function, the discrete Fourier transform is used:
\begin{equation}
  \label{ft_discrete}
  F_k = \sum^{N-1}_{n=0} f_ne^{-2 \pi i n k/N},
\end{equation}
where $f_n$ denotes the $n$-$th$ sample of $f$.
Fourier transform of a time series provides its spectrum over the entire period of time measured.
In order to detect changes in the signal spectrum over time, the short-time Fourier transform (STFT) is used.
STFT applies the Fourier transform to short overlapping segments of the time series.

Frequency domain analysis provides several advantages.
First, it provides a more compact and concise representation of most of the dominant signal components.
Second, it allows for the detection of attacks involving changing the frequency of regular operation modes, e.g., quickly starting and stopping the engine. Lastly, according to the uncertainty principle~\cite{de1967uncertainty}, functions localized in the time domain (e.g., a short spike) are spread across many frequencies, and functions that are concentrated in the frequency domain are spread across the time domain.
This means that slow attacks that usually evade time domain detection methods will stand out in frequency analysis, but short attacks will be difficult to detect using it.

\subsection{\label{subsec:adv_attacks}Adversarial Attacks on Machine Learning Models}
In this section, we provide a brief overview of adversarial attacks. A more complete review of related work in the context of this research is presented in Section \ref{sec:related}.
Adversarial data is specially crafted input samples that cause the algorithm to produce incorrect results at test time.
The field of adversarial learning in DNNs has gained a lot of interest since \cite{szegedy2013intriguing} showed that neural network-based classifiers can be tricked into mislabeling an image by changing a small number of pixels in a way that is imperceptible to the human eye.
Since then, successful adversarial attacks on neural networks have been demonstrated in  malware detection, speech classification, and other areas.

Adversarial machine learning attacks can be divided into \textbf{poisoning} attacks performed at training time and \textbf{evasion} attacks performed at test time.
In order to model adversarial attacks, we need to consider the attacker's goals and knowledge. 
According to \cite{biggio2018wild}, the goals are further subdivided into the desired violation (integrity, availability, or privacy) and specificity (targeting a set of inputs or all of them, as well as producing specific output or just any incorrect output).
In the context of anomaly detection, the attacker's goal might be to cause the system to classify an anomaly as benign (specific integrity attack) or to classify many benign samples as anomalous to decrease the trust in the detection results (indiscriminate availability attack).

Bigglio \etal \cite{biggio2018wild} define the attacker's knowledge in terms of the training data $\mathcal{D}$, the feature set  $\mathcal{X}$, the algorithm $f$ and its objective function $\mathcal{L}$, the training hyperparameters, and the detection parameters $w$ learned.
In the context of our research, $\mathcal{X}$ is the set of sensors' and actuators' states used to train the model, while $f$ and $\mathcal{L}$ represent the selected neural network architecture and its loss function.
Thus, the attacker's knowledge is represented by the components $(\mathcal{D}, \mathcal{X}, f, w)$.  
The worst-case perfect knowledge white-box scenario happens when all four components are known.
Gray-box attacks occur if at least one of the components is not known and cannot be reproduced.
For example, the attacker might know the feature set and the neural network type, but the network parameters and weights are not known.
In such cases, the attacker tries to create a surrogate model using training data sets relevant to the problem and transfers the attack created on the surrogate model onto the real one.
Black-box attacks are characterized by the lack of specific knowledge about any of the four components.
The attacker, however, knows that some model is used for the task at hand and can make educated guesses about the kind of features it uses to solve it.

\section{\label{sec:related}Related Work}
The area of anomaly and intrusion detection in ICSs has been widely studied.
Extensive surveys~\cite{mitchell2014survey, han2014intrusion, humayed2017cyber} and surveys of surveys~\cite{giraldo2017security} are devoted to the classification of research in this field.
In our review of related work, we focus on ICS anomalies and cyber attack detection using \textit{the physical state of the system} as measured by the sensors.
As noted in~\cite{giraldo2018survey}, the first step in physics-based detection is system state prediction.
By observing the deviation between the predicted and reported system state, a decision is made on whether an attack or anomaly occurred and how to score it.
Hence, one of the main ways to classify the research is by the prediction method used.
Auto-regressive (AR) models are used to predict the system state in~\cite{hadvziosmanovic2014through} and~\cite{mashima2012evaluating}. 
While popular in time series analysis, these models have limitations in multivariate systems, when the state of one observed variable is correlated with another.
In our research, we use DNNs that don't have these limitations.

Another popular way of modeling the system is rooted in the control theory and uses the subsystem model identification based on equation~\eqref{eq:lds} which describes a linear dynamical system:
\begin{equation}
\label{eq:lds}
\begin{array}{lcl} 
x_{k+1} & = & Ax_k + Bu_k + \epsilon_k \\
y_k & = & Cx_k + Du_k + e_k,
\end{array}
\end{equation}
where $x_k$ is the system state at time $k$, $u_k$ denotes the controller commands to actuators, $y_k$ are the sensor measurements, $\epsilon_k$ is perturbation noise, $e_k$ is sensor noise, and $A, B, C, D$ are matrices modeling the dynamics of the system. 
This approach has been used in previous studies, such as~\cite{mishra2017secure, murguia2016characterization, mo2015physical}.
The limitations of linear dynamical system modeling include the requirement for controller command measurement, a requirement which is not met in most datasets.
In addition, many attack scenarios involve altering PLC logic and do not violate system dynamics. 
For those reasons, we chose to use DNNs that are more flexible on both counts.

Specification-based system modeling can also be very effective, as shown by~\cite{mitchell2015behavior} and~\cite{mishra2019modeling}.
In~\cite{mitchell2015behavior}, the authors used behavior rules to specify the safe system state for medical Cyber Physical Systems (CPSs) and monitor deviation from these rules.
Distributed invariant-based mechanisms for smart grids are presented in \cite{rahman2016multi} and \cite{roth2018physical}.
In \cite{roth2018physical}, detection is based on observing the physical state of the shared system, detecting the power conservation invariants' violation and identifying the rogue component by the invariants' verification in its topological neighborhood.
While effective in rogue CPS controller identification, the solution proposed in \cite{roth2018physical} is very specific to smart grids, where the physical invariants are well-known and simple.
Rahman~\etal~\cite{rahman2016multi} used multiple computationally powerful agents that communicated with each other.
One of the main drawbacks of these approaches is their specificity - the solution should be tailored to the system and its operating conditions; in contrast, our approach is generic and requires no manual configuration.

PASAD is a novel approach to the problem presented in \cite{aoudi2018truth}.
PASAD is based on ideas from singular spectrum analysis and detects attacks in the signal subspace representing the deterministic part of the time series. 
The main idea of PASAD is to break the signal into subseries and find their noise-reduced representation by singular value decomposition. 
The principle difference between our detection approach and PASAD is the ability of our approach to detect anomalies in correlation among input features, while PASAD is limited to a single time series. 

In a recent competition on water distribution system cyber attack detection (the BATADAL - BATtle of the Attack Detection Algorithms \cite{taormina2018battle}), seven teams demonstrated their solutions on a simulated dataset.
The best results were shown by the authors of \cite{housh2018model}, who were able to  model the system precisely using MATLAB. 
The main limitation of this solution is its reliance on the need and ability to create a precise system model, both a non-generic and difficult task.
Another work that achieved a high score in the competition is \cite{abokifa2017detection} in which the authors proposed a three-layer method, where the first layer detects statistical anomalies, the second layer is a neural network aimed at finding contextual inconsistencies with normal operation, and the third layer uses principal component analysis (PCA) on all sensor data to classify the samples as normal or abnormal.
Our work differs from \cite{abokifa2017detection} in the following ways.
First, we study the efficiency of a single generic mechanism, as opposed to the multilevel system used by \cite{abokifa2017detection}.
Second, our solution evaluates types of neural networks not covered by \cite{abokifa2017detection}.
In addition, we study frequency domain anomaly detection and adversarial robustness.

Another relevant study from the BATADAL competition is~\cite{chandy2017detection}.
In addition to other detection mechanisms, the authors of~\cite{chandy2017detection} used VAEs to calculate the reconstruction probability of the data.
In our research, we found that VAEs are not very accurate in reconstructing time series data. 
Therefore, we suggest using simpler autoencoder models and demonstrate their effectiveness at this task.

Neural networks have been used in additional physics-based cyber attack detection research (\cite{goh2017anomaly, inoue2017anomaly, lin2018tabor}).
Unlike our work, these studies use more complex recurrent and graphical models and do not study the frequency domain.

Autoencoders have been used for anomaly and intrusion detection before \cite{mirsky2018kitsune,  sakurada2014anomaly}.
The differences between this work and~\cite{mirsky2018kitsune} are that in our research (1) AEs are applied to raw physical signals without statistical feature extraction, and (2) AEs are applied to the frequency domain.
We extend the research in \cite{sakurada2014anomaly} by applying AEs to cyber attack detection in time series, combining control, status and raw physical data, as well as applying AEs to the frequency domain. 
We also enhance the architecture of the network and present a feature selection method which improves network performance.

After our research was complete we discovered a recent publication by Taormina \etal that applied autoencoders to the BATADAL dataset \cite{taormina2018deep}.
The authors  demonstrated the effectiveness of AEs and were able to achieve an $F1$ score of 0.886.
Also, the authors showed how to obtain insight on the attack location from the model's predictions.
Our research was performed in parallel to \cite{taormina2018deep}, was not influenced by its findings, and differs from it in the following ways:
\begin{itemize}
    \item we study both 1D CNNs and AEs on three different datasets, two of which come from real-world testbeds,
    \item our AE architecture is different because it has been adopted to multivariate time series prediction, uses noise and an inflation layer; it also achieves a higher $F1$ score,
    \item we study frequency domain detection, and
    \item we present adversarial attacks on the proposed network and its robustness. 
\end{itemize}

Little attention has been given to adversarial attacks in the ICS context, and there are a number of differences between  our study and the work in the area where most of adversarial research has been done (image and sound processing):
\begin{itemize}
    \item most of the existing work is focused on supervised learning problems, while our research deals with a semi-supervised learning,
    \item most of the existing work is focused on classification tasks, while in our research we deal with prediction (regression),
    \item while in tasks such as image classification the output variable (picture class) is not part of the input, in our task the input and output features are same, and
    \item in our case, there are multiple constraints on the internal structure of the data, due to the laws of physics and PLC logic, that are not present in images.
\end{itemize}

A successful evasion attack framework on machine learning anomaly detection was demonstrated in \cite{ghafouri2018adversarial}.
The authors were able to bring a monitored reactor to a dangerous pressure level by manipulating sensor measurements in a way that was classified by both a linear regression and a feed-forward neural network as normal. 
In \cite{feng2017deep}, the authors used a GAN (generative adversarial network) in order to create stealthy attacks on an ICS, evading a baseline anomaly detector.
Most recently, Erba \etal \cite{erba2019real} showed how to create successful evasion attacks against an autoencoders-based detection mechanism.
The main difference of \cite{erba2019real} and our approach is the threat model chosen.
The authors of \cite{erba2019real} consider a very powerful attacker that can both generate arbitrary malicious inputs to the PLC and create fake traffic that is fed to the detector.
We consider a more constrained attacker that can only control the sensory data that is both seen by the PLC and the detector.
We argue that our threat model represents a more realistic scenario.

\section{\label{sec:datasets}Datasets}
\subsection{\label{subsec:swat}SWaT}
The Secure Water Treatment (SWaT) testbed was built at the Singapore University of Technology and Design. 
Although a detailed description of the testbed and dataset can be found in \cite{goh2016dataset}, we provide a brief description below.
\begin{figure}[ht]
\centering
\includegraphics[clip, trim=6cm 12.3cm 6cm 12cm]{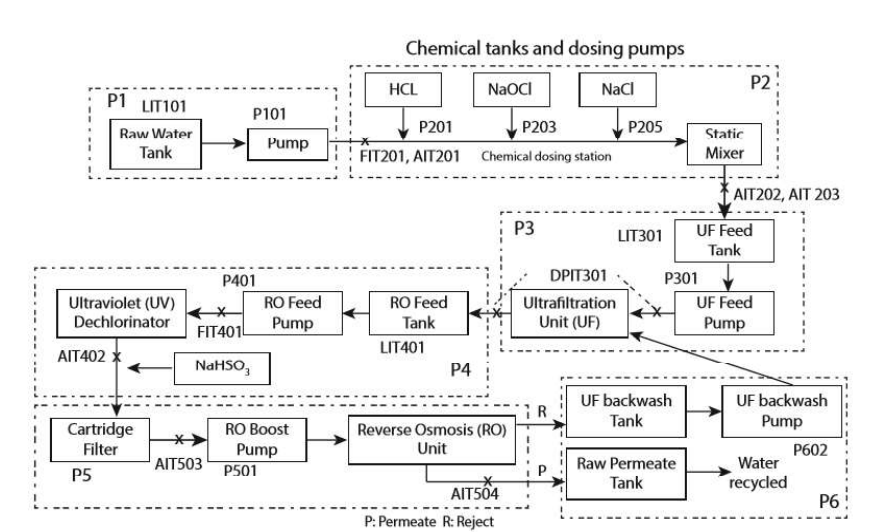}
\caption{SWaT testbed process overview \cite{goh2016dataset}.}
\label{fig:swat}
\end{figure}
The testbed is a scaled-down fully operational water treatment plant.
As shown in Figure~\ref{fig:swat}, the water goes through a six-stage process. 
Each stage is equipped with a number of sensors and actuators.
The sensors include flow meters, water level meters, and conductivity and acidity analyzers.
Water pumps, chemical dosing pumps, and valves that control inflow are the actuators.
The sensors and actuators of each stage are connected to the corresponding PLC, and the PLCs are connected to the SCADA system.

\begin{figure}[t!]
\centering
\includegraphics[clip, scale = 0.6, trim=0cm 0cm 0cm 0cm]{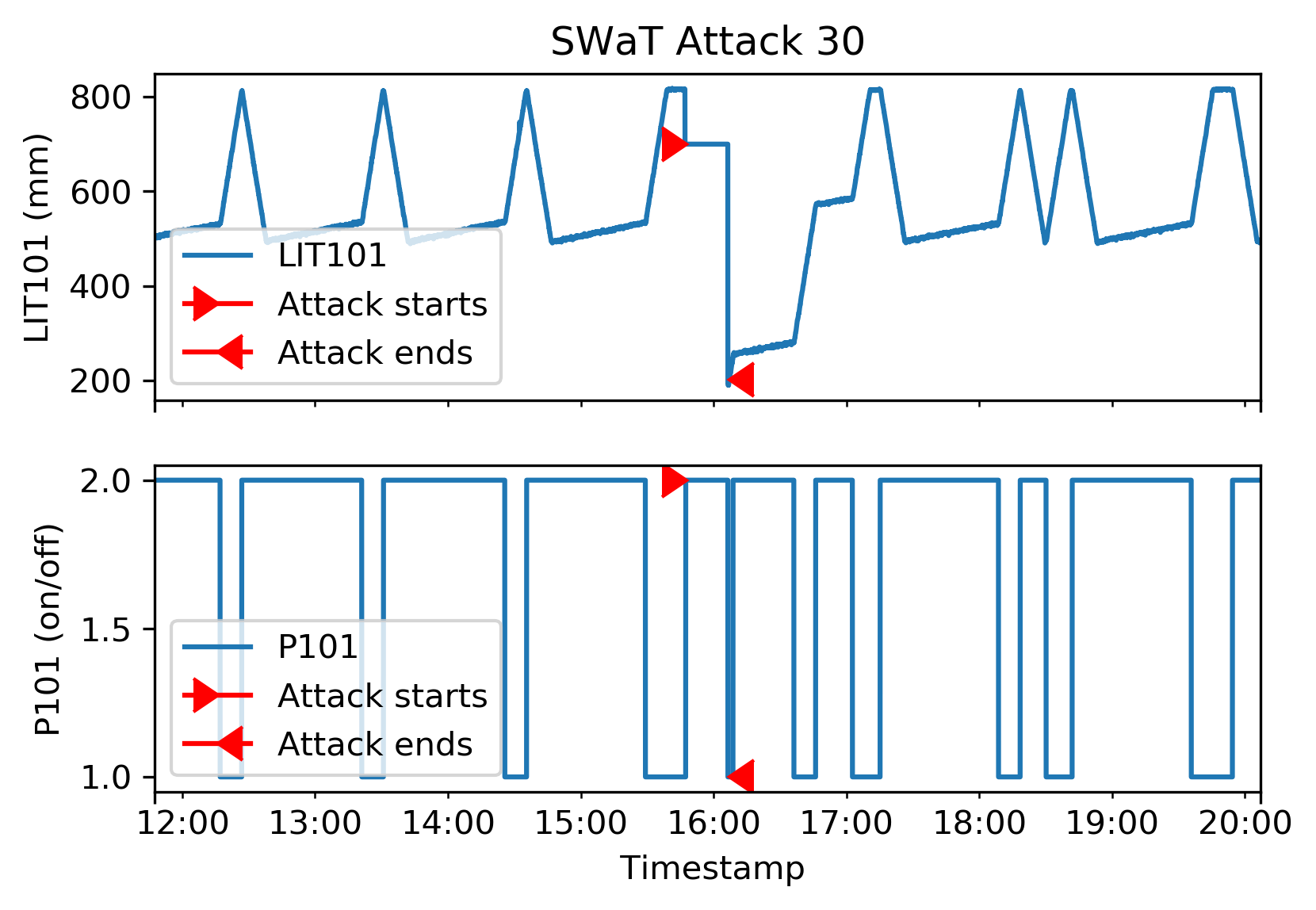}
\caption{Attack 30 on the LIT101 sensor. 
LIT101 measures the water level in the first tank. 
P101 pumps the water out of the tank to the second stage processing.
Note that after the attack is over, it takes a long time until the system returns to its normal production cycle. }
\label{fig:attack_30}
\end{figure}

The dataset contains seven days of recording under normal conditions and four days during which 36 attacks were conducted.
The entire dataset contains 946,722 records, labeled as either attack or normal, with 51 attributes corresponding to the sensor and actuator data.
The threat model used in the experiment is a system that has already been infected by attackers who spoof the system state to the PLCs causing  erroneous commands to be issued to the actuators, or override the PLC commands with malicious ones. 
A table containing a description and the timing of the attacks is provided in \cite{goh2016dataset}.
Each attack aims to achieve some physical effect on the system.
For example, attack 30 aims to cause underflow in the tank of the first stage.
For that purpose, the value of the water level sensor LIT101 is fixed at 700mm, while pump P101, which controls water outflow is kept open for 20 minutes.
Figure~\ref{fig:attack_30} presents the attack, its effects, and the time it takes the system to stabilize.
The attacks were usually not stealthy, i.e., when a command was issued to the actuator and the actuator changed the system state, the change was not hidden by the attackers.
\subsection{\label{subsec:batadal}BATADAL}
\begin{figure}[b!]
\centering
\includegraphics[clip, scale = 0.4, trim=0cm 0cm 0cm 0cm]{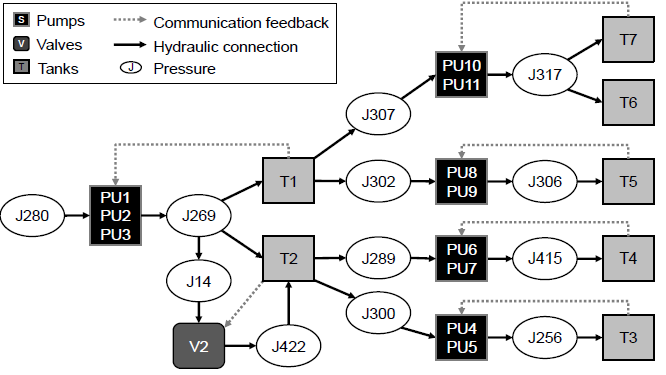}
\caption{Hierarchy of the water distribution system used in the BATADAL dataset~\cite{chandy2018cyberattack}.}
\label{fig:batadal}
\end{figure}

The BATADAL dataset represents a water distribution network comprised of seven storage tanks with eleven pumps and five valves, controlled by nine PLCs (see Figure~\ref{fig:batadal}).
The network was generated with epanetCPA \cite{taormina2017characterizing}, a MATLAB toolbox that allows for the injection of cyber attacks and simulates the response of the network to these attacks.

There are 43 variables representing the water tank levels, the flow and status of all of the pumps, as well as the inlet and pressure for the pumping stations and valves.
The training data simulates hourly measurements collected for 365 days, resulting in 8,761 records.
The test dataset contains 2,089 records (from 87 days of recording).
There are seven attacks present in the test data.
The attacks involved malicious actuator activation, PLC set point changes, and sensor measurement manipulation.
In addition, the attacks were concealed from the SCADA system by replacing the PLC-to-SCADA communication data with the data recorded at the same hour during normal operation.
\begin{figure}[t!]
\centering
\includegraphics[clip, scale = 0.6, trim=0cm 0cm 0cm 0cm]{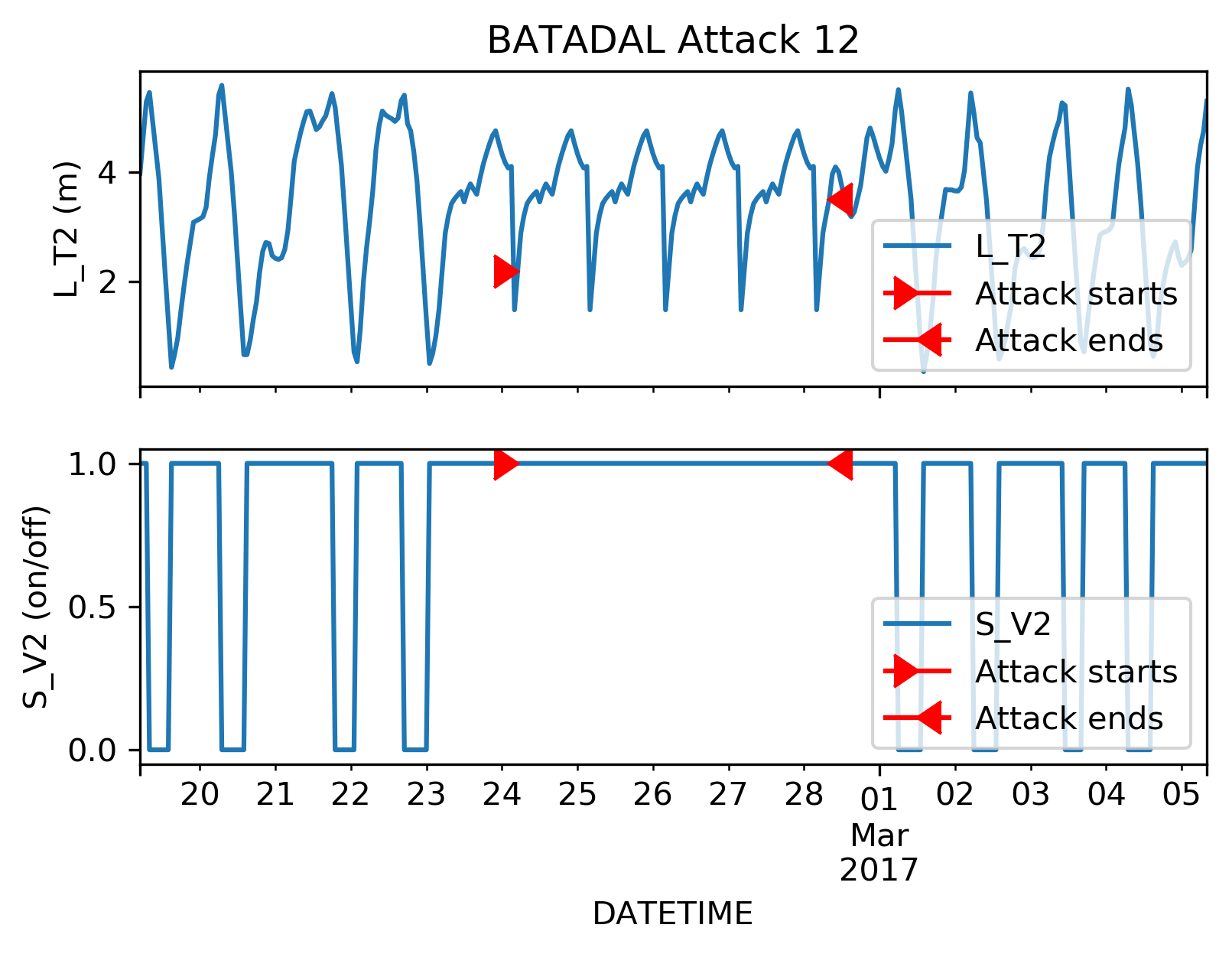}
\caption{Attack 12 on the L\_T2 sensor.}
\label{fig:batadal_attack_12}
\end{figure}
Figure \ref{fig:batadal_attack_12} illustrates attack 12.
The goal of this attack is to cause tank T2 to overflow.
The L\_T2 sensor's readings are altered to report lower levels and cause PLC3 to keep the valve V2 open.
At the same time, the traffic from PLC3 to SCADA is modified to replay previously recorded  values of L\_T2, as well as V2 flow and pressures. 
Figure \ref{fig:batadal_attack_12} shows that the status of the valve was not replayed, although the authors of \cite{taormina2018battle} reported that it was.
Also, one can see that immediately after the attack the system returns to its regular cycle.
This looks unrealistic, as the tank must be in an overflow state and it should take time to process the excess water it contains.
We estimate that these represent limitations of the simulation.

\subsection{\label{subsec:wadi}WADI}
Finding another high-quality real-world cyber physical dataset containing attacks is not an easy task. 
The best candidate we could find is WADI~\cite{ahmed2017wadi}, collected from a scaled-down water distribution testbed and built by the authors of SWaT. 
The testbed consists of a number of large water tanks that supply water to consumer tanks. The dataset contains 16 attacks whose goal is to stop the water supply to the consumer tanks. 
The attacks were conducted by opening valves and spoofing sensor readings, and were partially concealed. 
The dataset is significantly larger than the SWaT and BATADAL datasets, and contains 1,209,610 data points in the training set and 126 features. 
The WADI dataset was made public recently, and very few attack detection results utilizing this dataset have been published. 
In~\cite{mishra2019modeling}, the authors proposed an agent-based framework for CPS modeling and used it to detect attacks on the WADI dataset.
Unfortunately, the authors of~\cite{mishra2019modeling} did not publish the quantitative metrics of the detection results, only reporting that 12 of 16 attacks were detected.
In \cite{li2019mad}, the authors use LSTM-based generative adversarial networks (referred to as MAD-GANs in the paper) and show that they outperform other methods, such as PCA, K-nearest neighbors (KNN), and feature bagging (FB) on the SWaT and WADI datasets.

\section{\label{sec:method}Methodology}
\subsection{Data Analysis and Preprocessing}\label{sec:method_preproc}
\begin{figure}[b!]
\centering{\includegraphics[clip, trim=0cm 0cm 0cm 0cm, scale=0.22]{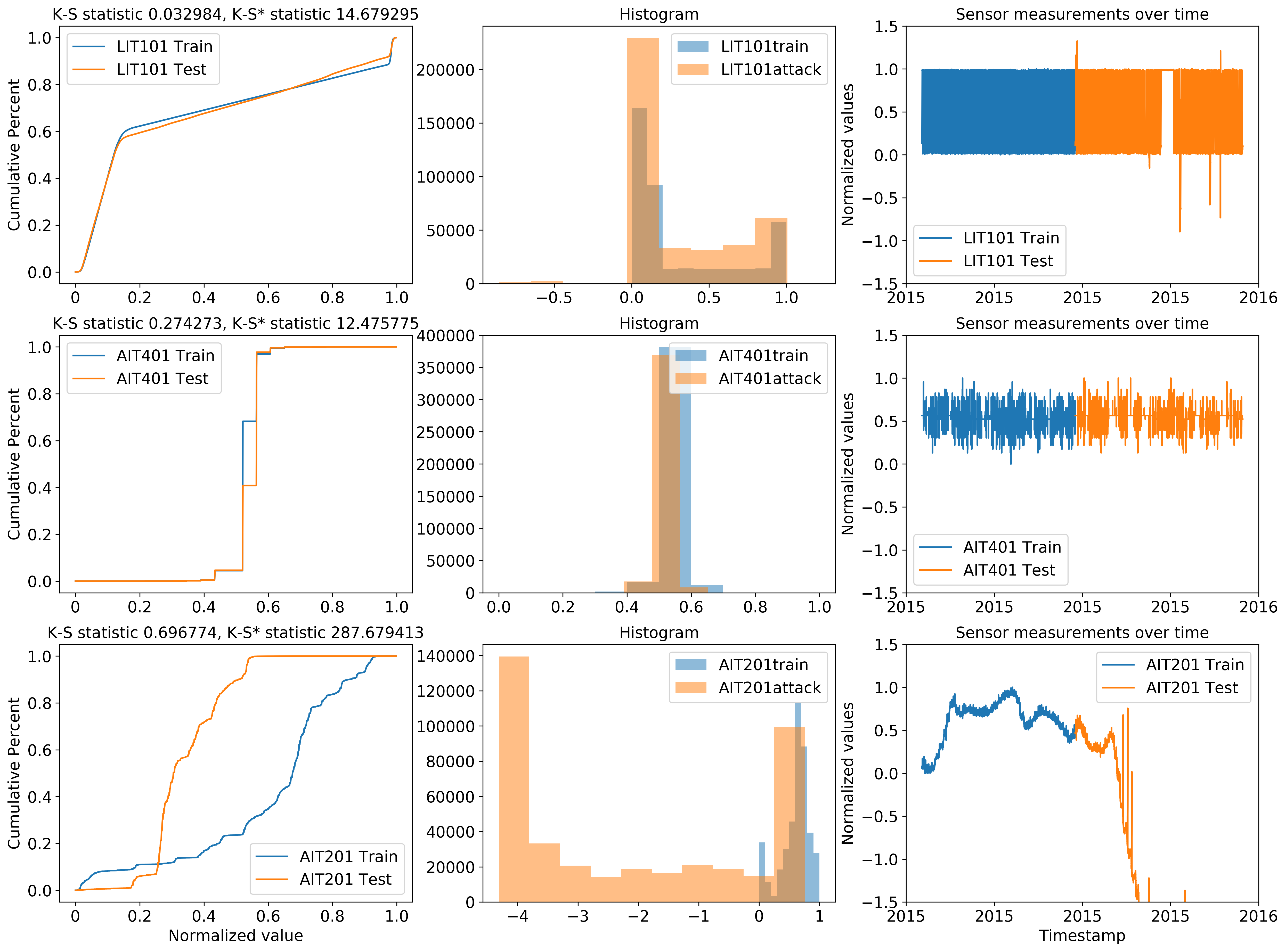}}
\caption{Feature statistic comparison. LIT101 has a very similar distribution in both the training and test data. AIT401 has a similar but slightly offset distribution. K-S has a high value, but K-S* correctly classifies the distributions as close. AIT201 has very different distributions.}
\label{fig:feat_stat}
\end{figure}

Our detection mechanism is based upon the ability to model and predict the system's behavior.
To fulfill this requirement, the following assumptions must hold: the training data must be representative of the test data.
More specifically, the training data should contain all of the (latent) states, and the transitions between them that appear in the test data.
In other time-series forecasting techniques, e.g., AR models and recurrent neural networks, there is a stronger requirement that the data needs to be stationary (i.e., maintains its probability distribution over time) or can be transformed into stationary \cite{goodfellow2016deep} form.
We found that a number of SWaT features do not have the same distribution in the training and test data (see Figure~\ref{fig:feat_stat}).
In order to obtain a quantitative measure of the similarity between the probability distributions of the training and test data, we used the Kolmogorov-Smirnov test (K-S test) \cite{chakravati1967handbook}. 
We chose the K-S test, because it is non-parametric and isn't based on any assumptions on the probability distributions tested. 
It also is more sensitive than comparing the mean and standard deviation or the t-test, both of which do not work well with multimodal and non-normal distributions.

The K-S test statistic for two distributions is the maximal difference between their empirical cumulative distribution functions (ECDF):
\begin{equation}
  \label{eq:KS-test}
    K\mhyphen S = \sup_{x}\left| F_1(x) - F_2(x)\right|,
\end{equation}
where $F_1$ and $F_2$ are ECDFs of the compared distributions.
They can be found as:
\begin{equation}
  \label{eq:CDF}
F_x = \frac{1}{n}\sum_{i=1}^{n}{I_{[-\infty,x]}(X_i)},
\end{equation}
where 
\[
    I_{[-\infty,x]}= 
\begin{cases}
    1,& \text{if } X_i<x\\
    0,              & \text{otherwise}.
\end{cases}
\]

The original K-S test is limited to fully specified distributions \cite{natrella2010nist}, however we found the slight modification described below useful as a concise metrics for filtering out features unsuitable for modeling.
Using the maximum as a statistic makes the K-S test extremely sensitive to small CDF differences when the distribution's mean is slightly offset on the x axis.
To increase the test's robustness, we used the area between the CDFs instead, which is calculated as:
\begin{equation}
  \label{eq:KS*-test}
    K\mhyphen S^{*} = \int_{x}\left| F_1(x) - F_2(x)\right| dx.
\end{equation}
Figure \ref{fig:feat_stat} illustrates three SWaT features, their values over time, histograms, and K-S and K-S* statistics.

\begin{figure}[b!]
\centering{\includegraphics[clip, trim=0cm 0cm 0cm 0.3cm, scale=0.26]{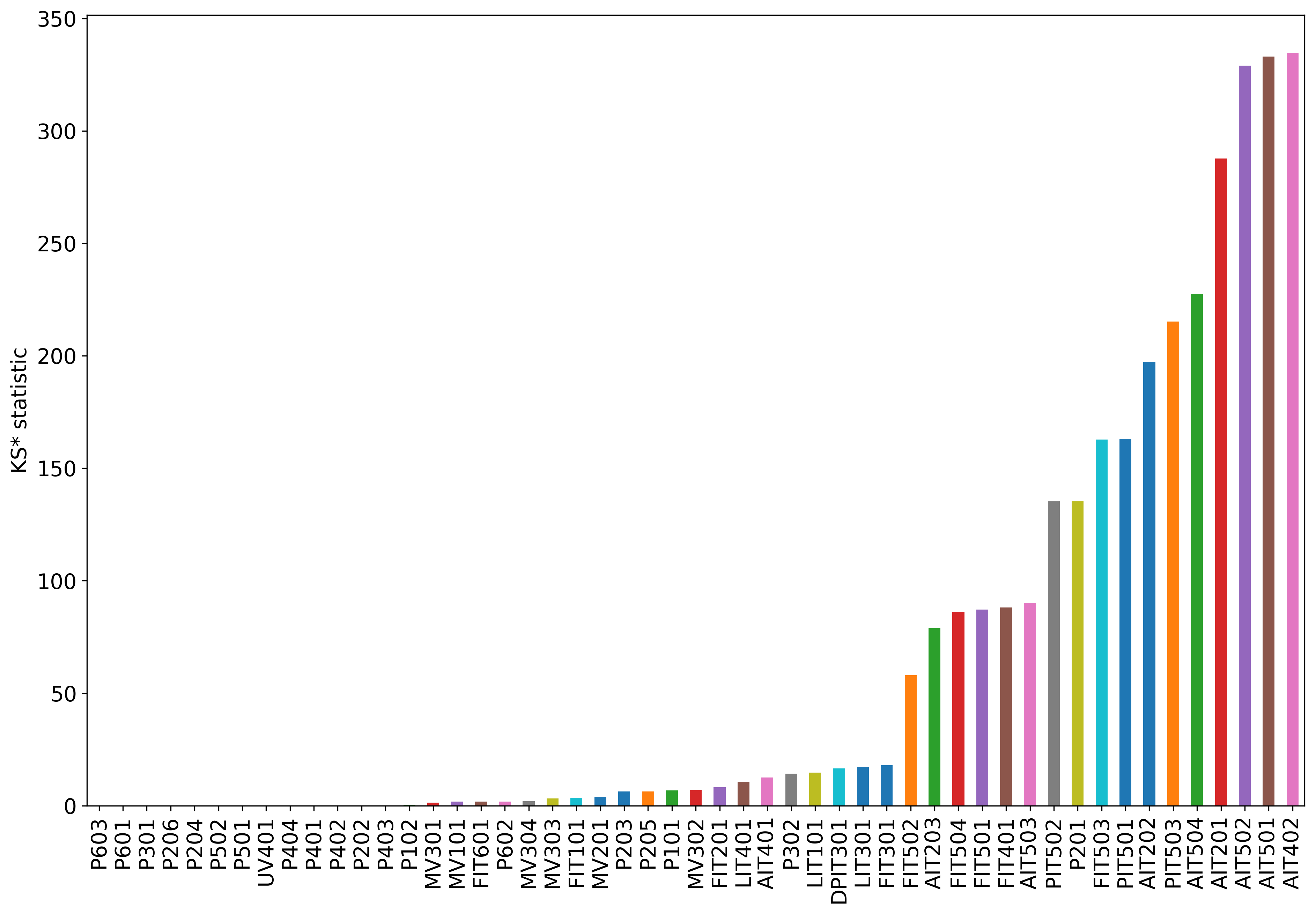}}
\caption{$K\mhyphen S^{*}$ statistic for the SWAT dataset.
A number of features differ significantly between the training and test sets.}
\label{fig:ks_swat}
\end{figure}

We calculated the $K\mhyphen S^{*}$ statistic for all SWaT and BATADAL features.
The features were normalized to (0,1) scale.
As Figure \ref{fig:ks_swat} shows, many of the SWaT features differ greatly between the training and test sets.
Such features would create a lot of false positive alarms and must be excluded from the modeling.
In addition to data normalization and feature statistic profiling, we subsampled the SWaT data at a five second rate. 
Subsampling provides a regularization mechanism which prevents overfitting and allows us to operate with a smaller amount of data.

As for the BATADAL dataset, all but one (P\_J280) of its features have very low K-S* metrics (10 or less).
This striking difference between the real-world and simulated data stresses the need to validate any findings in realistic setups.

For the WADI dataset, subsampling at a ten second rate was applied, and twelve unstable features were removed.

The feature selection step should be done prior to model training, hence requiring different treatment for different data availability scenarios:
\begin{itemize}
    \item if both training and test datasets are provided, the selection should be done based on both of them,
    \item if only the training dataset is available, the test can be done on two parts of it, e.g., comparing the statistics of the first half of the data to the second, and
    \item a periodic validation of the features' consistency can be performed during the test period, and the model should be retrained if significant changes are detected.
\end{itemize}

\subsection{Undercomplete Autoencoders Design}\label{subsec:method_ae}
After experimenting with multiple AE architectures, including LSTM-based AEs, variational AEs, and denoising AEs, we discovered that the best detection performance is achieved with the simplest undercomplete AEs.
We describe the selected AE architecture here, and a comparison of the results in provided in Section~\ref{sec:experiments_ae}.

\begin{figure}[ht]
\centerline{\includegraphics[clip, trim=0cm 0cm 0cm 0cm, scale=0.6]{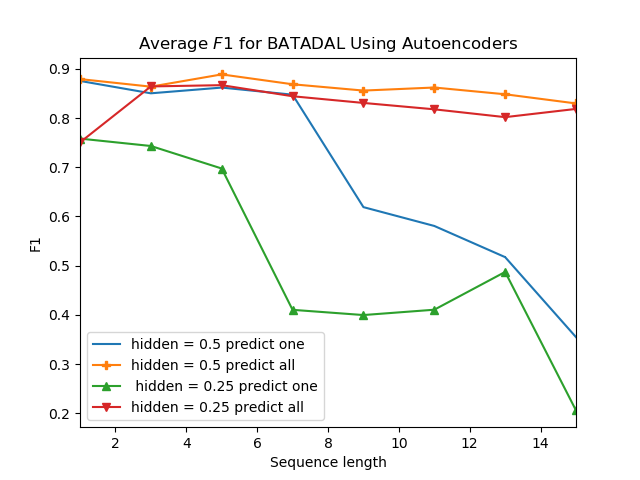}}
\caption{Improvement of detection score when measuring loss for the entire sequence in autoencoders.}
\label{fig:AE_all_vs_one}
\end{figure}

The best results were achieved using the AE network variant adapted for multivariate sequence reconstruction.
\begin{figure*}[t!]
\centering{\includegraphics[clip, trim=0cm 0cm 0cm 0cm, scale=0.5]{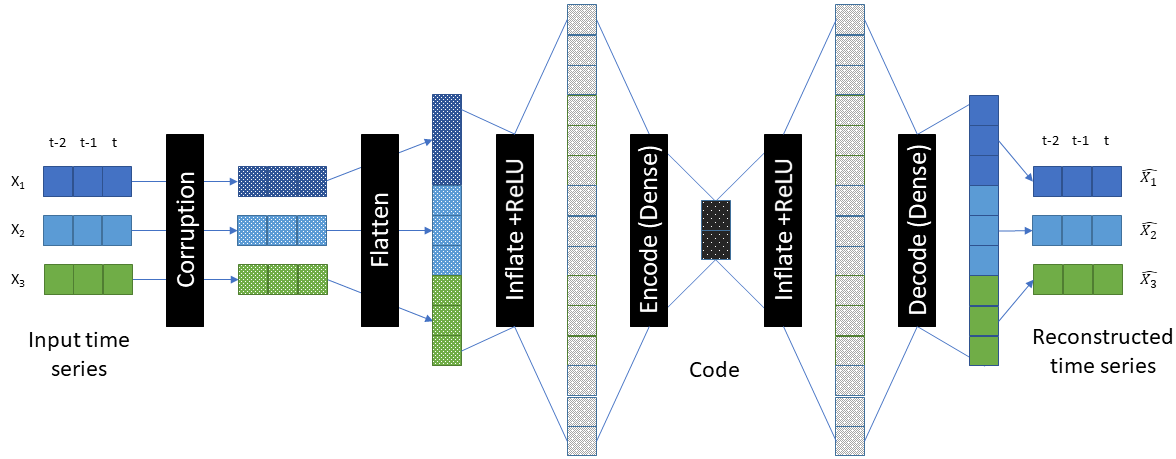}}
\caption{Autoencoders architecture used in this research.}
\label{fig:ae_arch}
\end{figure*}
The network design is as follows:
\begin{itemize}
    \item an optional corruption layer applying Gaussian noise to the input sequence,
    \item a fully connected layer with an ReLU or tanh activation function \textbf{inflating} the input; the purpose of this layer is to enlarge the hypothesis space,
    \item an encoding layer that flattens the input and produces its compact representation using a fraction of the input size; in our experiments the best results were achieved using the compact representation twice smaller than the input,
    \item a decoding layer reconstructing the original sequence from its compact representation.
\end{itemize}
This architecture can deal with time sequences of arbitrary length and is presented in Figure~\ref{fig:ae_arch}.
We conducted a comparison between the detection performance of a loss function based on the reconstruction error for the predicted data points only versus the reconstruction error for the entire sequence.
The latter produced much better results as illustrated in Figure~\ref{fig:AE_all_vs_one}.

\subsection{\label{subsec:method_pca}Principal Component Analysis}
Principal component analysis (PCA) transforms a set of variables to a set of values of linearly uncorrelated variables, orthogonal to each other.
The new variables are linear combinations of the original ones.
PCA is based on eigenvector analysis and is often used for reducing the dimensions of the data to a small number of principal components that represent the data's internal structure in the way best explaining the data variance.
We used PCA as a baseline algorithm to compare the performance of the neural networks.
The use of PCA for anomaly detection in ICS is not new; it was suggested as one of the detection layers in \cite{abokifa2017detection}, and \cite{sakurada2014anomaly} used it for detecting anomalies in a simulated Lorenz system and telemetry data.
In \cite{abokifa2017detection}, detection was conducted in \textbf{principal components subspace}.
Anomalies found in this subspace do not have a direct physical meaning in the original data feature space, and are hard to interpret and explain.

In our research, we used the approach outlined in  Algorithm \ref{alg_pca_transform}.
\begin{algorithm}[b!]
\caption{Predict $x_{test}$ given training data $x_{train}$ and using PCA.}
\label{alg_pca_transform}
\begin{algorithmic}[1]
\Function{pcaAnalysis}{$x_{train}, x_{test}, components$}
\State $pcaModel \gets PCA(components)$
\State $pcaModel.fit(x_{train})$
\State $x_{test}PCA \gets pcaModel.transform(x_{test})$ 
\State $\hat{x}_{test} \gets pcaModel.inverse\_transform(x_{test}PCA))$
\State \textbf{return} $\hat{x}_{test}$
\EndFunction
\end{algorithmic}
\end{algorithm}
Our analysis restores the prediction to the original feature space thus allowing for the natural application of the detection and explanation method we use for neural networks, as described below.
To distinguish this detection method from detection in the principal components' subspace, we refer to it as \textit{PCA-Reconstruction} in  Section~\ref{sec:experiments}.
We implemented an extension to the classic PCA analysis.
PCA usually operates on single time step vectors.
This allows for the detection of context inconsistencies between multiple features but is less powerful in the detection of time-related inconsistencies in a single feature.
To compensate for this deficiency, we implemented a \textit{windowed-PCA} algorithm, which breaks the data into time windows of a given width, performs the analysis depicted in Algorithm \ref{alg_pca_transform} on vectors containing multiple data points, and then restores the window predictions into the original signal shape.
We later discovered that this idea was described in \cite{ku1995disturbance}, where it is called Dynamical PCA.
Two variants of this \textit{windowed-PCA} algorithm were implemented: with overlapping and non-overlapping windows.
The standard PCA algorithm implementation from the Python scikit-learn package with the number of components equal to half of the modeled features was used as a basis for our experiments.

\subsection{\label{subsec:method_freq}Frequency Domain Transformation}
\begin{figure}[t!]
\centering{\includegraphics[clip, trim=0cm 0cm 0cm 0.0cm]{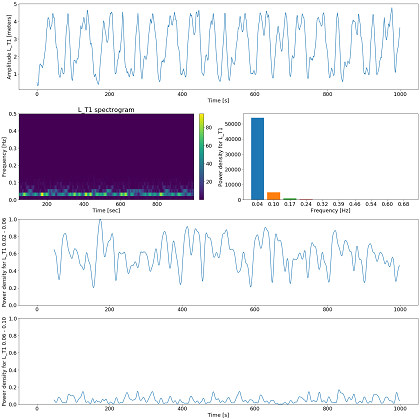}}
\caption{Frequency transformation of the L\_T1 feature. The first row depicts the raw signal. The second row represents the STFT spectrogram on the left and the power density distribution in the spectrum bands on the right. The last two rows are the frequency domain features - the power density in the first two bands. }
\label{fig:lt1_freq_transform}
\end{figure}

In order to explore the usefulness of frequency domain attack detection we first had to transform the signals from the time domain to the frequency domain. 
The following method was used to create signal representation in the frequency domain (outlined in Algorithm \ref{alg_freq_transform} and illustrated in Figure ~\ref{fig:lt1_freq_transform}).

\begin{algorithm}[b!]
\caption{Transform signal $s$ into frequency domain representation}
\label{alg_freq_transform}
\begin{algorithmic}[1]
\Function{frequencyAnalysis}{$s, sampling\_period$}\Comment{Find the dominant frequency of the signal and its period}
\State $N\gets len(s)$
\State $freq \gets DFFT(s)$\Comment{The discrete FFT of the signal}
\State $magn \gets abs(freq)[:N // 2] * 1 / N$\Comment{The magnitudes of the real part of the FFT}
\State $freq\_magn \gets listOf(freq, magn)$
\State $sorted\_freq\gets decreasedOrderSort(freq\_magn)$
\If{$sorted\_freq[0][0]$}\Comment{Ignore the constant component if it has the most energy}
\State $fundamental\_freq \gets sorted\_freq[0][0]$
\Else
\State $fundamental\_freq \gets sorted\_freq[1][0]$
\EndIf
\State $period \gets (1.0/fundamental\_freq)/sampling\_period$
\State \textbf{return} $(fundamental\_freq, period)$
\EndFunction

\Function{frequencyTransform} {$s, rate, ratio, b\_num$}\Comment{Represent  signal as energy in the most dominant frequency bands.}
\State $all\_freq\_bins \gets 10$
\State $(f\_freq, period) \gets frequencyAnalysis(s, rate)$
\State $STFT\_window \gets period * ratio$ 
\State $freqs, Sx \gets spectrogram(s, rate, STFT\_window)$
\State $bands \gets linspace(0, len(freqs), all\_freq\_bins + 1)$
\For {$i = 0$ to $all\_freq\_bins$} 
\State $bands\_energy_i \gets \sum^{bands_{i+1}}_{i=bands_i} Sx_i$
\EndFor
\State $dominant\_bands \gets decreasedOrderSort(bands\_energy)[:b\_num]$
\State \textbf{return} $dominant\_bands$

\EndFunction

\end{algorithmic}
\end{algorithm}

\begin{enumerate}
\item Determine the dominant frequency of each signal (the frequency with the most energy) using the discrete fast Fourier transform (DFFT)(function frequencyAnalysis, lines 1-12).
\item Determine the window for the short-time Fourier transform (STFT) based on the dominant frequency period. It was found that the optimal window is between one and two periods of the dominant frequency (line 16).
\item Transform the signals into their frequency representation. 
\begin{enumerate}
    \item Split the entire signal into overlapping windows. 
    \item Perform STFT for each window (these two items are presented by the call to spectrogram at line 17).
    \item Binarize the entire spectrum of STFT into a number of bins. Calculate the total energy of the signal in each bin.
    \item Pick a small number of bins with the most energy (lines 18-22). The energy values will represent the feature in the frequency domain for the corresponding time window. We found that two or three bins were sufficient for representing the features. It is also possible to calculate the number of bins based on the ratio of the total energy they contain (e.g., at least 90\%).
\end{enumerate}
\item Apply the chosen neural network model (1D CNN or AE) to the frequency domain representation. As each feature is represented separately, the ability to locate the attack is maintained.
\end{enumerate}
\subsection{\label{subsec:method_adv}Adversarial Threat Model and Robustness Analysis}
As this research was the first study of the proposed detection method's robustness to adversarial attacks, we assumed the worst-case scenario - a white-box attacker that knows everything about the model used for the detection, including its weights.
We consider an attacker that is trying to perform a specific integrity attack, namely to cause a physical level change in the system's behavior while staying undetected by the monitoring anomaly detection system; the attacker can influence the values of sensors sent to the PLC but does not have complete control of the network, in either the remote or the control segment.
Such an attack scenario is very common,  especially in ICSs with sensors distributed over a large area and that send their data to a PLC residing in a physically protected and monitored center.
In this setup, an adversary can replace the original sensor with a malicious one, reprogram the sensor, change its calibration, influence the sensor externally, or just send false data to the PLC over the cable/wireless connection.
We argue that this setup is much more practical than an attacker controlling the internal network of the remote segment or even the network of the control segment, the scenario studied in \cite{erba2019real}.
In our threat model the attacker's input is processed by \textbf{both the PLC and the detection system}, hence the attacker's goal is to produce the input that will at the same time:
\begin{enumerate}
    \item cause the intended physical impact on the system, and
    \item be close enough to the prediction of the detector to stay under the detection threshold.
\end{enumerate}
Other characteristics of our threat model include:
\begin{itemize}
    \item the attacker can change multiple sensors,
    \item the attacker can prepare the attacks offline; we argue that if the system is characterized by periodic behavior, the attacker can choose the moment of the attack and precompute the system state in our perfect knowledge threat model, and
    \item the values of the sensors after the attack should be within or close to the valid range of the sensor values (e.g. an on/off sensor can't accept any other value).
\end{itemize}

As we did not possess access to the testbeds, we based our study on adversarial manipulation of the SWaT dataset attacks.
While this method cannot replace testing with a real system, it can provide an approximation of the ability to produce the desired adversarial inputs.
The two main limitations of such data-only research are:
\begin{enumerate}
    \item it is limited to the attacks already present in the data set; even if none of them can't be concealed by an adversarial input there may be other attacks that have this capability,
    \item there is no way of testing the physical effect of an adversarial sample found analytically on the real system. 
\end{enumerate}

The adversarial robustness study was conducted as follows.
First, eighteen attacks from the SWaT dataset caused by spoofed sensors were selected.
For each attack we performed gradient-based search for the adversarial input as outlined in Algorithm~\ref{alg_adversarial_input}.
This algorithm is adapted to a sequence prediction model that processes the input using subsequences of length $l$, which we assume is also known to the attacker.

\begin{algorithm}[ht]
\caption{Find $x_{adv}$ given a trained model $\mathcal{M}$, test data with attack $x_{att}$, sub-sequence length $l$, a detection threshold $\tau$, acceptable noise level $\epsilon$, input constraints $\phi$.} 
\label{alg_adversarial_input}
\begin{algorithmic}[1]
\Function{findSubSeqAdvInput}{$\mathcal{M}$, $x_{att}$, $l$, $\tau$, $\epsilon$, $\phi$}
\State $\mathcal{M'} \gets \mathcal{M} + \nabla_{x}{J(\theta,x,y)}$\Comment{Add to the model graph the cost function gradient calculation $J$ given the model parameters $\theta$, input $x$ and correct output $y$ with respect to the input $x$. $ADV\_LR$ is the adversarial learning rate}
\State $x_{adv} \gets [] $ \Comment{Initialize as empty}
\For {$ss \gets nextSubSequence(x_{att}, l)$} 
\State $noise \gets zeros\_like(ss)$
\State $advIt \gets 0$
\While {$advIt < MAX\_ADV\_ITERATIONS$}
    \State $noisy\_input \gets ss + noise$
    \State $noisy\_input \gets enforceConstraints(noisy\_input, \phi)$
    \State $model\_residue, grad \gets runModel(\mathcal{M'} $
    \If{$model\_residue < \tau$}
        break
    \EndIf
    \State $step \gets ADV\_LR * max(abs(grad))$
    \State $noise \gets noise - step * grad$ \Comment{Update the noise}
    \State $noise \gets clip(noise, \epsilon)$\Comment{Make sure the noise does not pass the acceptable level}
    \State $advIt \gets advIt + 1$
\EndWhile
\State $x_{adv}.append(noisy\_input)$
\EndFor

\State \textbf{return} $x_{adv}$
\EndFunction
\end{algorithmic}
\end{algorithm}

\begin{figure}[t!]
\centering{\includegraphics[clip, trim=0cm 0cm 0cm 0cm, scale=0.5]{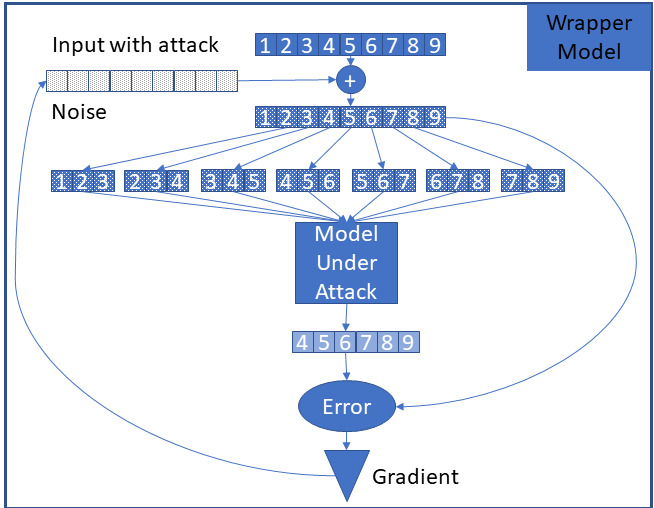}}
\caption{A wrapper model for adversarial learning. The underlying model under attack uses sequences of three time points to predict the subsequent time point.}
\label{fig:wrapper_model}
\end{figure}

However,  Algorithm~\ref{alg_adversarial_input} finds adversarial variants for individual subsequences, and does not consider the following constraints of the way neural networks are used in our method:
\begin{itemize}
    \item each data point is used in $l$ subsequences, where $l$ is the subsequence length;  
    \item as the previous data points are used to predict the next one, perturbing a data point at time $t$ will require changes to earlier data points so that the prediction at time $t$  will be close enough to the desired value; these changes will need to propagate back in time,
    \item if data gradients are used as enrichment features, as described in \cite{kravchik2018detecting}, the adversarial gradient calculation should consider these enrichment features as well, even though they are not part of the trained system model. 
\end{itemize}
In order to cope with these constraints, we needed to create a \textbf{wrapper model ($\mathcal{WM}$}) for the original model $\mathcal{M}$. 
$\mathcal{WM}$ represents the processing of all the input including enrichment feature generation and subsequence generation as one graph, allowing for full gradient propagation from the model's prediction to all of the original input, not just a specific subsequence. 
Figure~\ref{fig:wrapper_model} illustrates a wrapper model for a case when three time steps are used to predict a single subsequent time step, without considering additional feature enrichment.
The final algorithm used for adversarial input generation follows the logic of Algorithm~\ref{alg_adversarial_input}, with the following modifications:
\begin{enumerate}
    \item a wrapper model is created, instead of performing simple addition of the gradient (line 2),
    \item as the wrapper model optimizes all of the input, there is no need in for the loop in line 4,
    \item we added adaptive an learning rate update to accelerate optimization.
\end{enumerate}

\subsection{\label{subsec:method}Anomaly Detection and Scoring Method}
The anomaly detection method used in this research is based on the one we used in \cite{kravchik2018detecting}, however in the current study we extend it in a number of ways which are elaborated upon below.
We trained a neural network until its training error reached the desired value (usually less than 0.1).
The network is used to predict the future values of the data features based on previous values.
Thus, the model performs the function
\begin{equation}
  \label{model_func}
  (\hat{y}_{h+n}, \hat{y}_{h+n+1}, \ldots, \hat{y}_{h+n+m}) = f(y_{n-1-l}, \ldots, y_{n-1})
\end{equation}
where $y_i$ is a feature vector at time $i$, $\hat{y}_i$ is the estimation of the feature vector, $l$ and $m$ represent the input and output sequence length respectively, and $h$ is the prediction horizon.
We generalized the method to allow the prediction of arbitrary length sequences in the future with a specified horizon, e.g., predicting 256 time steps starting with the fifth time step from the last input time.
The residuals' vectors are calculated as:
\begin{equation}
  \label{residuals}
	\vec{r_t} = \left| \vec{y_t} - \vec{\hat{y_t}} \right|.
\end{equation}
The residuals are used to trigger the anomaly alert in one of two ways.
In the first, the residuals are normalized by dividing them by the maximal per feature residuals for the training data, and the maximum of the normalized residuals is compared to a threshold $\tau$:
\begin{equation}
    \label{eq:norm_residual}
	\vec{R_t} = \frac{\vec{r_t}}{\max{\vec{r}}}. 
\end{equation}

In order to prevent false alarms on short-term deviations, we require that the residual exceed the threshold for at least a specified duration of time window $w$.
Thus, an anomaly alert $A_i$ at time $i$ is determined by: 
\begin{equation}
    \label{alert}
	A_i = \displaystyle \prod_{t=i-w}^i \max \vec{R_t} > \tau. 
\end{equation}

The hyperparameters $\tau$ and $w$ are determined by setting a maximal accepted false alarm rate for the validation data and finding the solution to:
\begin{equation}
  \label{eq:threshold_window}
	(\tau, w) = \argmin_{\tau,w} \omega_\tau \cdot \omega_w \{(\tau,w) \mid \left|A(\tau,w) \right| \leq FP_{max}\},
\end{equation}
where $\omega_\tau$ and $\omega_w$ are weights of the threshold and the window correspondingly, $A(\tau,w)$ is the set of attack alerts detected with the specific threshold and window values, and $FP_{max}$ is the maximal allowed number of false alarms in the validation data.
In other words, we are looking for the hyperparameter values that don't produce more than the permitted number of false alerts, while minimizing the product of their weights. The weight of a hyperparameter is proportional to its index in the argument space. For example, if the possible window value space is {5, 10, 15, 20}, the corresponding weights will be {0.25, 0.5, 0.75, 1}. Using weights allows us to normalize the contribution of both hyperparameters regardless of their absolute values.

The second way to detect the attacks differs from the first one by normalizing the residuals using their mean and standard deviation (by feature) and is described in~\cite{kravchik2018detecting}. 
In this research, we found that in the case of the SWaT dataset, using the residuals' mean and standard deviation based on the test data produced better results than using the statistics based on the training data. 
Updating threshold statistics with test data, a common practice in online anomaly detection, compensates for data drift. 
This finding hinted at the presence of data drift in the SWaT dataset, and it was indeed detected and dealt with, as we described in Section~\ref{sec:method_preproc}.

In order to produce results \textbf{comparable} with previous research, we used the \textbf{same performance metrics} as other works using the corresponding dataset. 
For SWaT and WADI, the metrics are precision, recall and $F1$ and they are calculated based on log record labels contained in the dataset.

In the BATADAL competition, the score was calculated as a weighted sum:
\begin{equation}
  \label{eq:batadal_score}
    S = \gamma \cdot S_{TTD} + (1 - \gamma) \cdot S_{CLF}
\end{equation}
where $S_{TDD}$ is the time-to-detection score, $S_{CLF}$ is the classification score, and $\gamma$ determines the relative importance of the two scores and is set to 0.5. 
The details of the calculation of both scores are based on the log record labels and are described in \cite{taormina2018battle}.
To summarize, our extensions to the anomaly detection method used in \cite{kravchik2018detecting} are:
\begin{itemize}
    \item generalization of the prediction, allowing arbitrary length sequence prediction and arbitrary prediction horizon,
    \item addition of max-based method for threshold detection, and
    \item formalization of the hyperparameters criteria.
\end{itemize}

\section{\label{sec:experiments}Experiments and Results}

\subsection{Evaluating 1D CNN Performance with BATADAL Dataset}\label{sec:experiments_cnn_batadal}
\begin{figure}[ht]
\centerline{\includegraphics[clip, trim=0cm 0cm 0cm 0cm, scale=0.6]{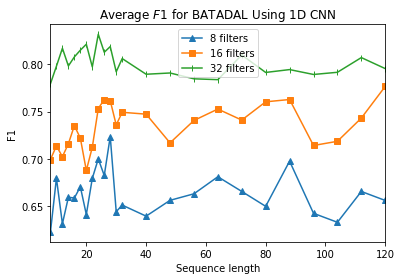}}
\caption{Influence of 1D CNN filters on detection score for BATADAL. Results for 8 layers model.}
\label{fig:BATADAL_1D_CNN}
\end{figure}

To answer our first research question, we validated the effectiveness of 1D CNNs with the BATADAL dataset.
We modeled all features, except for P\_J280, due to its high K-S* value.
Multiple hyperparameter configurations were tested, both using grid search and genetic algorithms \cite{goldberg1988genetic}.
The best score presented in Table~\ref{tab:perf_batadal} was achieved with an eight-layer 1D CNN using  a sequence length of 18 data points. 
From the frequency domain analysis, which is described later, we learned that the period of the dominant frequency for BATADAL is 24 hours (data points).
Therefore, 18 points represent a trade-off between capturing enough historical information and including too much data which causes less precise predictions.
Figure~\ref{fig:BATADAL_1D_CNN} illustrates the influence of hyper parameters on the detection score. 

\begin{table}[t!]
\begin{threeparttable}
\caption{Comparative performance of neural networks on the BATADAL dataset\tnote{1}}
\begin{center}
\begin{tabular}{|c|c|c|c|c|c|}
\hline
Method & Attacks & Score & $S_{TDD}$ & $S_{CLF}$ & $F1_{rec}$\\
\hline
Abokifa et al. \cite{abokifa2017detection} & 7 & 0.949 & 0.958 & 0.940 & 0.88\\
\hline
Chandy et al.\cite{chandy2017detection} & 7 & 0.802 & 0.835 & 0.768 & 0.538\\
\hline
\hline
1D CNN & 7 (1 fp) & 0.894 & 0.915 & 0.873 & 0.833\\
\hline
AE & 7 & 0.926 & 0.925 & 0.927 & 0.919\\
\hline
VAE & 7 (3 fp) & 0.882 & 0.919 & 0.846 & 0.783\\
\hline
PCA-Reconstruction & 7 & 0.898 & 0.898 & 0.898 & 0.875\\
\hline
\textbf{AE Frequency} & 7 & 0.969 & 0.980 & 0.959  & 0.937\\
\hline
\end{tabular}
\label{tab:perf_batadal}
\begin{tablenotes}
\small
\item[1] We omitted the results of  \cite{housh2018model}, who achieved the score of 0.97 and $F1$ of 0.97, as their approach is based on reconstructing the simulation parameters and is not applicable to real-world cases.
\end{tablenotes}
\end{center}
\end{threeparttable}
\end{table}
As Table~\ref{tab:perf_batadal} shows, the 1D CNN detected all of the attacks and achieved high scores, however it did not achieve the performance of the best BATADAL competitors.
Some of the attacks were not detected due to the attack concealment techniques used in BATADAL. 
Therefore, we conclude that while 1D CNN networks are indeed effective in detecting cyber attacks in CPSs, there is room for improvement in terms of precision, recall, and timeliness of detection.
\subsection{Undercomplete Autoencoders}\label{sec:experiments_ae}
As VAEs were used in a related work~(\cite{chandy2017detection}), we explored multiple VAE configurations, using both grid search and genetic algorithms (the best configuration results are presented in Table~\ref{tab:vae_ae_comparison}), and discovered that the generative nature of VAEs causes less precise predictions and lower recall.
\begin{table}[ht]
\begin{center}
\begin{threeparttable}
\caption{VAE and AE average $F1$ comparison for BATADAL.\tnote{1}}
\label{tab:vae_ae_comparison}
\begin{tabular}{|c|c|c|c|c|}
 \hline
  Length & 1 & 3 & 5 & 7\\
 \hline
 VAE & 0.745 & 0.690 & 0.719 & 0.773\\
 \hline
 AE & 0.879 & 0.864 & 0.889 & 0.869\\
 \hline
 \end{tabular}
 \begin{tablenotes}
\small
\item[1] Tests performed with the code size ratio of 0.5 and a single layer. 
\end{tablenotes}
\end{threeparttable}
\end{center}
 \end{table}

This VAE behavior is consistent with the results obtained in image generation, as reported in \cite{goodfellow2016deep}.
As Table~\ref{tab:perf_batadal} shows, VAEs obtained a lower score than simple non-generative AEs.

As shown in Table~\ref{tab:perf_batadal}, our simple AE network produced better detection scores than the 1D CNN and approached the best results of the competition winners; its attack location capabilities were also better than these of the 1D CNN.
As a point of comparison, we also used PCA with the same number of components used with the AEs.
As expected, the AEs performed better than PCA, as they are able to capture non-linear dependencies between features \cite{goodfellow2016deep}.
We were surprised to discover that our method of PCA-based anomaly detection showed excellent results, falling not far behind the AEs.
The result presented in Table~\ref{tab:perf_batadal} is for the PCA-Reconstruction algorithm; in contrast, windowed-PCA didn't show any improvement for BATADAL (unlike for SWaT and WADI) as shown in Table~\ref{tab:pca_f1}.
\begin{table}[b!]
\caption{PCA detection $F1$ scores for different sequence lengths.} \label{tab:pca_f1}
\centering
\begin{tabular}{|c|c|c|c|}
\hline
Length&SWaT&BATADAL&WADI\\
\hline
1&0.8172&0.8747&0.54\\
\hline
2&0.8312&0.8255&0.552\\
\hline
3&0.84&0.703&0.562\\
\hline
4&0.8505&0.6948&0.550\\
\hline
5&0.8648&0.6279&0.653\\
\hline
6&0.8652&0.6855&0.669\\
\hline
7&0.8788&0.6003&0.67\\
\hline
8&0.8552&0.5663&0.683\\
\hline
9&0.854&0.5962&0.656\\
\hline
10&0.8762&0.4791&0.664\\
\hline
\end{tabular}
\end{table}

To verify the effectiveness of our AE- and PCA-based detection in a more realistic setup, we applied both to the SWaT dataset.
\begin{table}
\caption{SWaT attack detection performance comparison.} \label{tab:swat_perf_comparison}
\centering
\begin{tabular}{|c|c|c|c|}
\hline
Method & Precision & Recall & F1\\
\hline
DNN\cite{inoue2017anomaly} & 0.983 & 0.678 & 0.803\\
\hline
SVM\cite{inoue2017anomaly} & 0.925 & 0.699 & 0.796\\
\hline
TABOR\cite{lin2018tabor} & 0.862 & 0.788 & 0.823\\
\hline
1D CNN\cite{kravchik2018detecting} & 0.968 & 0.791 & 0.871\\
\hline
\hline
PCA-Reconstruction & 0.885 & 0.759 & 0.817\\
\hline
Windowed-PCA (window=7) & 0.92 & 0.841 & 0.879\\
\hline
AE & 0.890 & 0.803 & 0.844\\
\hline
AE Frequency & 0.924 &  0.827 & 0.873\\
\hline
\end{tabular}
\end{table}
As shown in Table \ref{tab:swat_perf_comparison}, the AEs obtained a high F1 score, comparable to the best results achieved using a 1D CNN in \cite{kravchik2018detecting}.
In addition, AE-based networks are smaller (for short sequences) and faster to train, as shown in Table~\ref{tab:times_sizes_comparison}.
\begin{table*}[ht]
\centering
\begin{threeparttable}
\caption{Training time and model size comparison for BATADAL.\tnote{1}}
\label{tab:times_sizes_comparison}
\begin{tabular}{|c|c|c|c|c|c|c|c|c|c|c|c|}
 \hline
  &\multicolumn{4}{|c|}{AE}&\multicolumn{3}{|c|}{1D CNN}&\multicolumn{4}{|c|}{PCA} \\
 \hline
  Seq. length &  1 & 3 & 5 & 18 & 18 & 18 & 18 & 1 & 3 & 5 & 18\\
 \hline
 Layers &  1 & 1 & 1 & 1 & 4 & 8 & 12 & 1 & 3 & 5 & 18\\
 \hline
1 training epoch time, s &  0.268&0.288&0.306&0.459&0.641&0.878&1.761&0.061&0.100&0.207&0.439\\
 \hline
Model Size, Kb & 67 & 587 & 1624 & 20957 & 697 & 3689 & 50417 & 6 & 15 & 24 & 81\\
 \hline
 \end{tabular}
\begin{tablenotes}
\small
\item[1] Both AEs and 1D CNN used a three-fold inflation layer. AEs did not use inflation in decoding. CNNs used 32 filters. 
\end{tablenotes}
 \end{threeparttable} 
\end{table*}

Again, we were very surprised by the excellent performance of both simple and windowed-PCA.
The most likely reason for this success is that in the SWaT dataset, many relations between features are linear, and PCA can capture them.
As PCA has an analytic solution that does not require iterative optimization, its training is much faster then the discussed neural networks (see Table~\ref{tab:times_sizes_comparison}).
This answers our second research question - AEs are a lightweight and effective alternative neural network architecture that can be used for anomaly and cyber attack detection in CPSs.
In addition, PCA provides a simple alternative that can be sufficient in many real-world setups.

\subsection{Attack Detection Explainability} \label{sec:experiments_interp}
\begin{figure}[t!]
\centering{\includegraphics[clip, trim=2cm 0cm 2cm 0cm, scale=0.25]{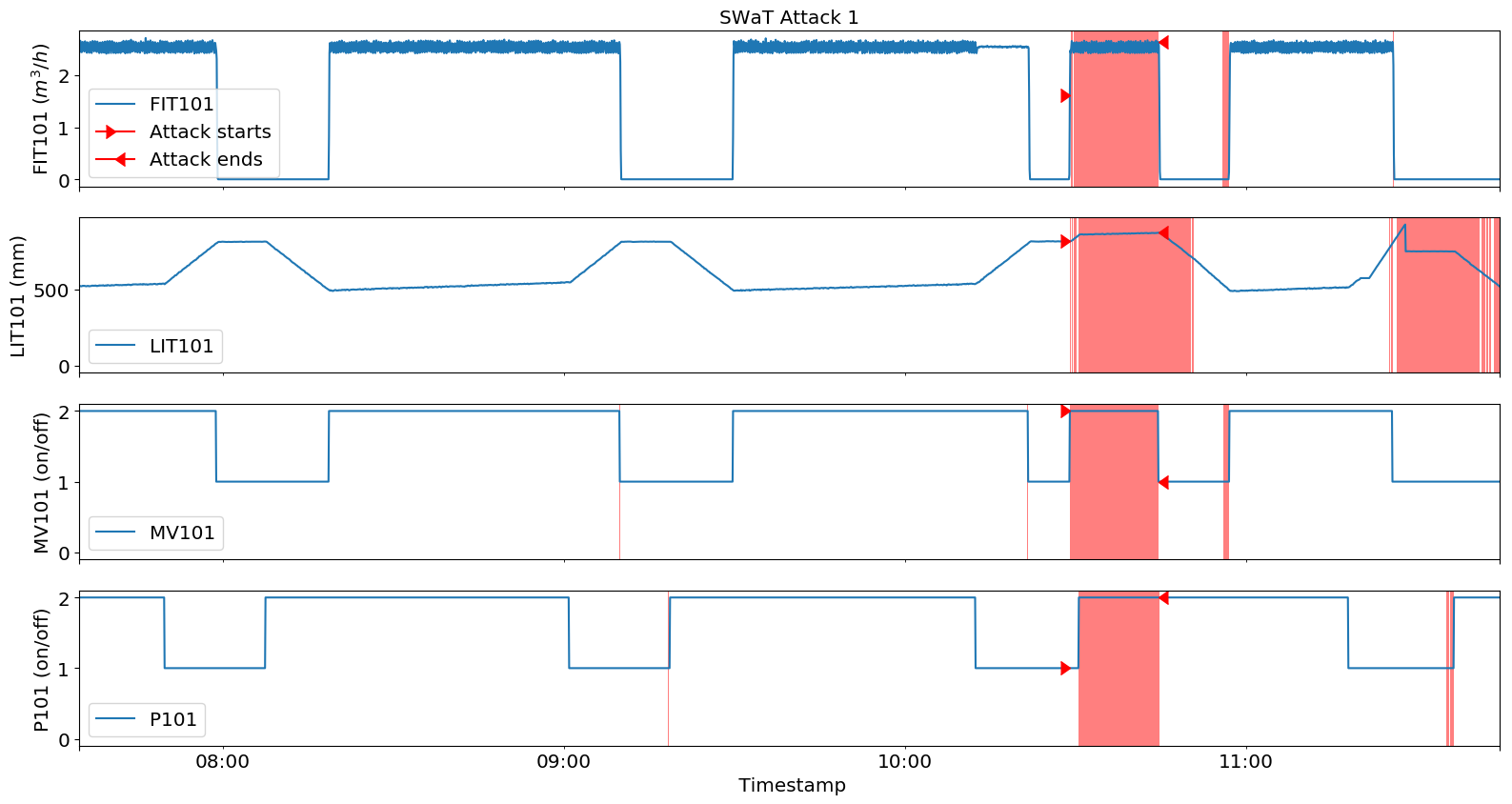}}
\caption{SWaT attack 1 location interpretation. The areas highlighted in red indicate the detected abnormal feature behavior. The attack opened the MV101 valve, letting more water into an already full tank. The model detects other related features as being abnormal too, e.g., water flow into a full tank (indicated by FIT101) and abnormally high water level (indicated by LIT101). Attacks 2 and 3, which were carried out soon after the first attack can be seen as well. Y-axes on all graphs represent normalized sensor values.}
\label{fig:attack1_interpretability}
\end{figure}

Once an attack has been detected, the ability to localize the attack is very important.
Using a neural network to model each feature in the monitored system allows us to assess which sensors and actuators were involved in the attack.
The attack indicator for a feature $i$ at a time $t$ is the corresponding residual $r^i_t$ bypassing the threshold $\tau$.
Analyzing the 1D CNN attack's location detection we observed the advantages of the combined feature modeling over modeling the features separately.
When each feature is modeled separately, the model often makes a prediction based on the recent past and thus is mainly useful for detecting abrupt non-characteristic changes of the feature.
To counter this effect, we increased the prediction horizon, so that recent past values become less useful.
This resulted in the discovery of more attacks as well as in more false positives.
On the other hand, when we modeled a number of features related to a single PLC or a number of related PLCs together, 1D CNN models capture dependencies between them.
This results is more complete and accurate attack detection, both in terms of time and location.
We also observed that spoofing a single feature might trigger behavior changes of multiple features, resulting in \textbf{all of them} being considered anomalous, as shown in Figure~\ref{fig:attack1_interpretability}.

Table \ref{tab:p1_swat_attacks_location} summarizes attack location detection for the attacks on the first stage of the SWaT testbed.
\begin{table}[b!]
\begin{threeparttable}
\caption{Attack location detection for the attacks in the first stage of the SWaT testbed\tnote{1}}
\begin{center}
\begin{tabular}{|c|m{3cm}|m{3cm}|}
\hline
Attack & Attack Description & Attacked Features Detected  \\
\hline
1 & Open MV101 to cause overflow & FIT101, LIT101, \textbf{MV101}, P101\\
\hline
2 & Turn on P102 to cause a burst pipe& FIT101, LIT101, \textbf{P102}\\
\hline
3 & Increase LIT101 to cause underflow & \textbf{LIT101}\\
\hline
21 & Keep MV101 on; set value of LIT101 at 700 mm & \textbf{LIT101}, \textbf{MV101}, P101\\
\hline
26 & Turn P101 on continuously to cause underflow & FIT101, LIT101, MV101, P102\\
\hline
28 & Close P302, causing the outflow to stop& LIT101, P101\\
\hline
30 & Turn P101 and MV101 on; set value of LIT-101 at 700 mm & \textbf{LIT101}, P102\\
\hline
33 & Set LIT101 to a high value to cause underflow & FIT101,\textbf{LIT101}, MV101\\
\hline
34 & Turn P101 off & \textbf{P101}, P102\\
\hline
36 & Set LIT101 to a low value to cause overflow & \textbf{LIT101}, P101\\
\hline
\end{tabular}
\label{tab:p1_swat_attacks_location}
\begin{tablenotes}
\small
\item[1] The features directly attacked appear in bold font. 
\end{tablenotes}
\end{center}
\end{threeparttable}
\end{table}

As Table~\ref{tab:p1_swat_attacks_location} shows, the 1D CNN can almost always identify the feature that was attacked directly, and can also locate the related features influenced by the attack.
The attack location detection in the BATADAL dataset is summarized in Table \ref{tab:batadal_attacks_location}.
In the BATADAL dataset, the attacks were concealed by replaying previously recorded valid data, but the network was able to detect them by detecting anomalies in the dependent features that were not replayed.
Regarding the AEs, as Table~\ref{tab:batadal_attacks_location} shows, in most of the cases, AEs were able to pinpoint the attacked features despite the concealment of the attack.

\begin{table}[b!]
\begin{threeparttable}
\caption{Attack location detection for BATADAL\tnote{1} \tnote{2}}
\begin{center}
\begin{tabular}{|m{0.7cm}|m{2.1cm}|m{2.1cm}|m{2.1cm}|}
\hline
Attack & Attack Description & Attacked Features Detected by 1D CNN & Attacked Features Detected by AE\\
\hline
8 & Change the L\_T3 thresholds to cause underflow & P\_J300, P\_J256 & \textbf{L\_T3}, L\_T7,...\\
\hline
9 & Fake L\_T2 readings to cause overflow & P\_J300, P\_J289\, P\_J422 & \textbf{L\_T2}, P\_J300, P\_J422\\
\hline
10 & Turn the pump PU3 on & F\_PU1, P\_J269 & L\_T1,  F\_PU2, P\_J269 ...\\
\hline
11 & Turn the pump PU3 on & L\_T1, F\_PU1, P\_J269, ... & L\_T1, F\_PU2, P\_J269, ...\\
\hline
12 & Fake L\_T2 readings to cause overflow & P\_J300, P\_J289 & P\_J300, P\_J289, P\_J422\\
\hline
13 & Change the L\_T7 thresholds & P\_J302, P\_J307 & \textbf{L\_T7}, P\_J302, P\_J307\\
\hline
14 & Alter the T4 signal to cause overflow in T6 & P\_J300, P\_J289, P\_J422 & \textbf{L\_T4}, P\_J300\\
\hline
\end{tabular}
\label{tab:batadal_attacks_location}
\begin{tablenotes}
\small
\item[1] The features directly attacked appear in bold font. 
\item[2] More features were detected as anomalies; only the most strongly indicated ones are listed.
\end{tablenotes}
\end{center}
\end{threeparttable}
\end{table}
To answer our third research question, our modeling method locates the attacked features, provided they are not replayed. 
However, we found that the attack may trigger a reaction in many features.
In that case our method will report \textbf{all features} influenced by the attack, without distinguishing the original cause from its consequences.
By comparing the list of features reported to be attacked (Table \ref{tab:batadal_attacks_location}) to the system architecture (Figure \ref{fig:batadal}) it is easy to see that the detected features are related to the \textbf{attack area}.
The same observation is true for SWaT.
So our method, while not being able to point out the attack point precisely in a general case, is able to locate the attack area, which provides significant value to the operator.
Moreover, we argue that given only historian-based data, it is \textbf{unfeasible in a general case to determine the root cause of the attack}.
Our claim is due to the fact that PLCs are hard real-time computers that should complete their scan cycle (reading the inputs, running the logic, and updating the outputs) in \textbf{milliseconds}.
Due to a much larger time granularity  of the historian records (one second for SWaT and WADI, one hour for BATADAL), our detection algorithm sees both the attack and its consequences in the same data point, and the \textbf{causality is lost}.
In order to be able to pinpoint the original attack point, we need to augment our dataset, possibly with the network data, and this item will be a topic of future research.

\subsection{Frequency Domain Detection} \label{sec:experiments_freq_dom}
Our fourth research question seeks to explore the usefulness of frequency domain attack detection.
Although the time and frequency domain represent the same information, the compactness of periodic signal representation in the frequency domain could help in detecting anomalies.
We used an AE network for both SWaT and BATADAL frequency domain detection.
The network consisted of one to three fully connected layers followed by an encoder and decoder(see Table~\ref{tab:freq_batadal_f1} for selected results).
\begin{table}[t]
\caption{Average $F1$ scores for BATADAL using AEs and frequency domain.} \label{tab:freq_batadal_f1}
\centering
\begin{tabular}{|c|c|c|c|c|}
\hline
Window & Step & Frequency Bands & Layers & F1\\
\hline
24 & 2 & 3 & 1 & 0.882\\
\hline
36 & 2 & 3 & 1 & 0.887\\
\hline
36 & 2 & 3 & 2 & 0.897\\
\hline
64 & 2 & 3 & 3 & 0.765\\
\hline
128 & 8 & 3 & 2 & 0.631\\
\hline
\end{tabular}
\end{table}

For BATADAL, we were able to match the score of the best previously published result \cite{housh2018model} (see Table~\ref{tab:perf_batadal}), however the detector of \cite{housh2018model} is built for the specific BATADAL configuration, while our architecture is generic.

For SWaT, we first conducted a statistical analysis of the frequency domain representation  and removed those features that differed significantly between the training and test data.
On the remaining data we were able to obtain an $F1$ score of 0.873, which is slightly better than the previously published results.
While these results are very encouraging and suggest further study and validation, we discovered one limitation of frequency domain detection.
In order to be able to transform the data into the frequency representation, we use windows of at least one period of the dominant frequency.
The consequence of this is a lack in the ability to distinguish between short attacks that quickly follow one another in the same window. 
Although in reality this might be a mild concern, in the SWaT dataset many short attacks occur in succession. 
Our method usually detects them as one long attack, which reduces the precision metrics. Thus, in general, the answer to our fourth research question is positive: \textbf{frequency domain analysis can contribute to attack detection, but it has its limitations.}

\subsection{Additional Validation with the WADI Dataset} \label{sec:experiments_wadi}
\begin{table}[b]
\begin{center}
\begin{threeparttable}
\caption{Average $F1$ detection scores for WADI.\tnote{1}}
\label{tab:wadi_ae_f1}
\begin{tabular}{|c|c|c|c|c|}
 \hline
  Length & 1 & 3 & 5 & 7\\
 \hline
 AE & 0.691 & 0.618 & 0.691 & 0.732\\
 \hline
 \end{tabular}
 \begin{tablenotes}
\small
\item[1] Tests performed with the code size ratio of 0.5 and a single layer. 
\end{tablenotes}
\end{threeparttable}
\end{center}
 \end{table}
With the WADI dataset we were able to achieve substantially better results than those reported in \cite{li2019mad}, as shown in Table~\ref{tab:wadi_performance}.
The unstable feature removal improved our ability to detect attacks in general, as our PCA result is significantly higher than the result reported in \cite{li2019mad}.
In this case, the windowed-PCA algorithm was able to improve the results of the PCA-Reconstruction.
For the 1D CNN, we modeled each PLC of WADI separately and then merged the detection.
A 1D CNN model with eight layers and sequences of 16 data points successfully detected 14 attacks and outperformed both PCA and MAD-GAN.
We should stress that the 16 data points belong to the data subsampled at a 1/10 rate and thus represent 160 original data points.
However, the best results in our experiments were achieved by autoencoders, using the AE model with sequences of length 7 (see Table~\ref{tab:wadi_ae_f1} for partial results presentation and  Table~\ref{tab:wadi_performance} for cross-algorithms best results comparison).

Unfortunately, the WADI dataset did not appear to be suitable for frequency domain analysis. 
Only 44 of 127 features had a clear dominant frequency, and the frequency was very low (with a period of 1440 minutes or 24 hours). 
Such very long periods result in poor resolution in detecting short attacks (attacks in the WADI dataset are about ten minutes long). 
Other features did not have any clear periodicity.
We consulted with the dataset authors who indicated that the production cycle of WADI is driven by consumer demand and these demands patterns changed hourly.
Thus WADI represents ICSs without a stable production cycle and therefore frequency domain analysis is not applicable to them.
To summarize, \textbf{we successfully validated our detection methods} on the WADI dataset. 
The absolute performance results of attack detection for the WADI dataset was lower than for SWaT and BATADAL.
Our communication with the WADI creators revealed that in addition to the irregular consumer patterns, there were more issues preventing precise detection, such as faulty sensors, sensors with high noise, and more.
Thus, WADI probably represents a more realistic case of the real world anomaly detection systems need to cope with. 
\begin{table}[b!]
\begin{center}
\begin{threeparttable}
\caption{Comparative performance of attack detection for the WADI dataset}
\begin{tabular}{|c|c|c|c|}
\hline
Method & Precision & Recall & F1\\
\hline
PCA\tnote{1} & 0.3953 & 0.0563 & 0.10 \\
\hline
KNN\tnote{1} & 0.0776 & 0.0775 & 0.08 \\
\hline
FB\tnote{1} & 0.086 & 0.086 & 0.09 \\
\hline
EGAN\tnote{1} & 0.1133 & 0.3784 & 0.17 \\
\hline
MAD-GAN\tnote{1} & 0.4144 & 0.3392 & 0.37 \\
\hline
\hline
PCA-Reconstruction & 0.763 & 0.571 & 0.653 \\
\hline
Windowed-PCA (window=4) & 0.807 & 0.593 &	0.683 \\
\hline
1D CNN & 0.697 & \textbf{0.731} & 0.714 \\
\hline
\textbf{AE} & \textbf{0.834} & 0.681 & \textbf{0.750} \\
\hline
\end{tabular}
\label{tab:wadi_performance}
\begin{tablenotes}
\small
\item[1] As reported in \cite{li2019mad}. 
\end{tablenotes}
\end{threeparttable}
\end{center}
\end{table}

\subsection{Adversarial Robustness of the Proposed Method} \label{sec:experiments_adversarial}
\begin{figure}[t!]
\centering{\includegraphics[clip, trim=0cm 0cm 0cm 0cm, scale=0.2]{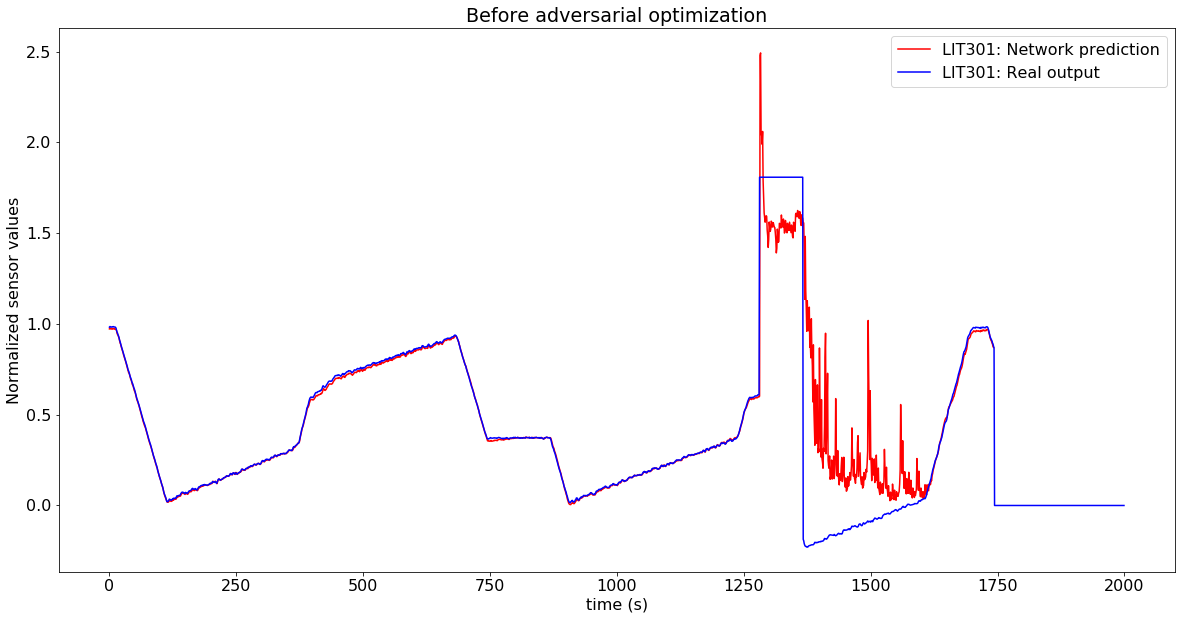}}
\caption{A 1D CNN model prediction for SWaT attack 7. During the attack and its recovery, the prediction is very different from the observed value.}
\label{fig:swat_attack_7_before_adv}
\end{figure}

First, the ability to create adversarial examples on a model of a single feature was tested.
In order to consider the worst-case scenario, no constraints were set on the allowed adversarial noise.
The experiments show that our wrapper model-based method is indeed capable of creating adversarial examples that cause the desired malicious physical effect and are not detected by the 1D CNN model.
\begin{figure}[t!]
\centering{\includegraphics[clip, trim=0cm 0cm 0cm 0cm, scale=0.2]{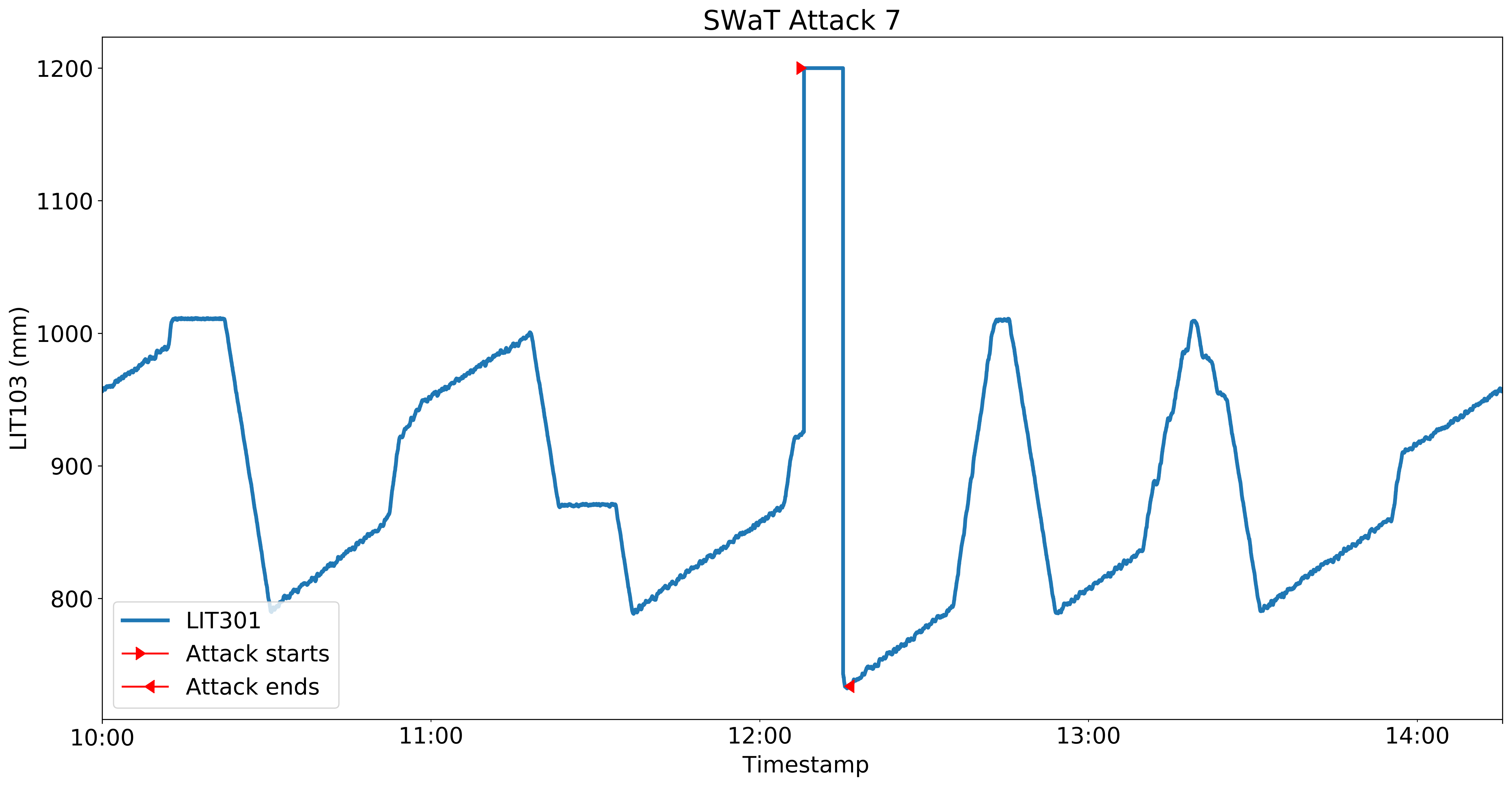}}
\caption{SWaT attack 7. LIT301 is spoofed to cause underflow.}
\label{fig:swat_attack_7}
\end{figure}
Figure~\ref{fig:swat_attack_7} illustrates attack 7 from the SWaT dataset. 
The measurement of the water level sensor LIT301 is spoofed to be much higher, causing underflow.
When an 1D CNN model created for LIT301 was used to detect anomalies in the relevant time period, it produced the prediction shown in Figure~\ref{fig:swat_attack_7_before_adv}.

\begin{figure}[t!]
\centering{\includegraphics[clip, trim=0cm 0cm 0cm 0cm, scale=0.2]{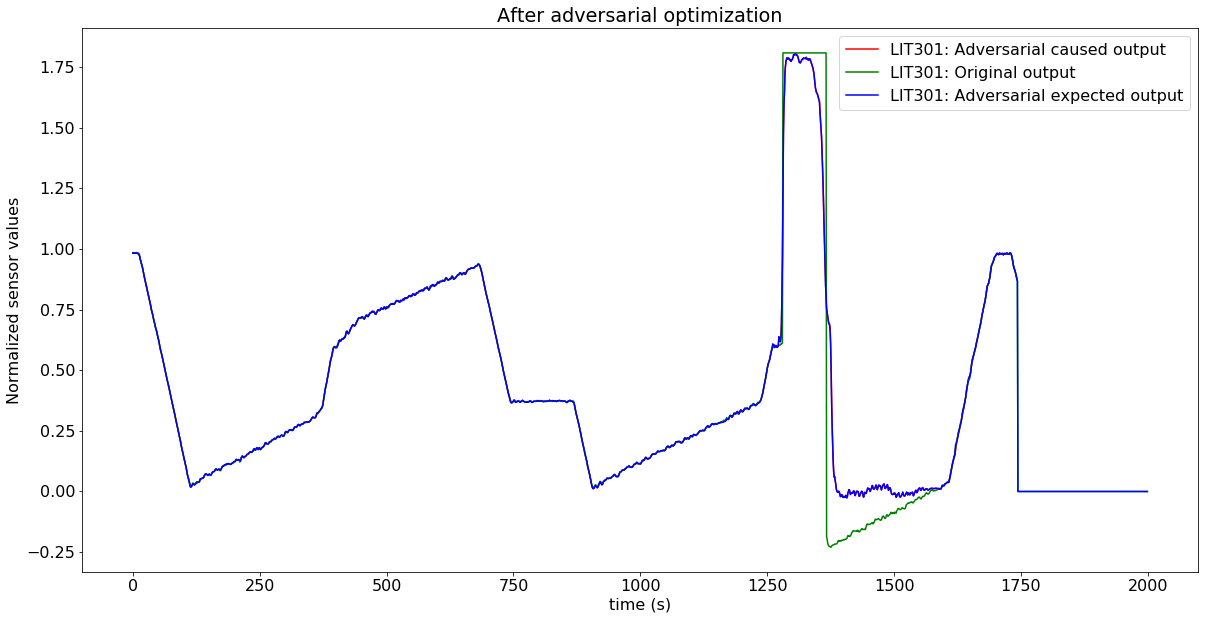}}
\caption{A 1D CNN model prediction for SWaT attack 7 after adversarial input optimization. The adversarial input is expected to cause underflow and is undetected by the model}.
\label{fig:swat_attack_7_after_adv}
\end{figure}
\begin{figure}[t!]
\centering{\includegraphics[clip, trim=0cm 0cm 0cm 0cm, scale=0.2]{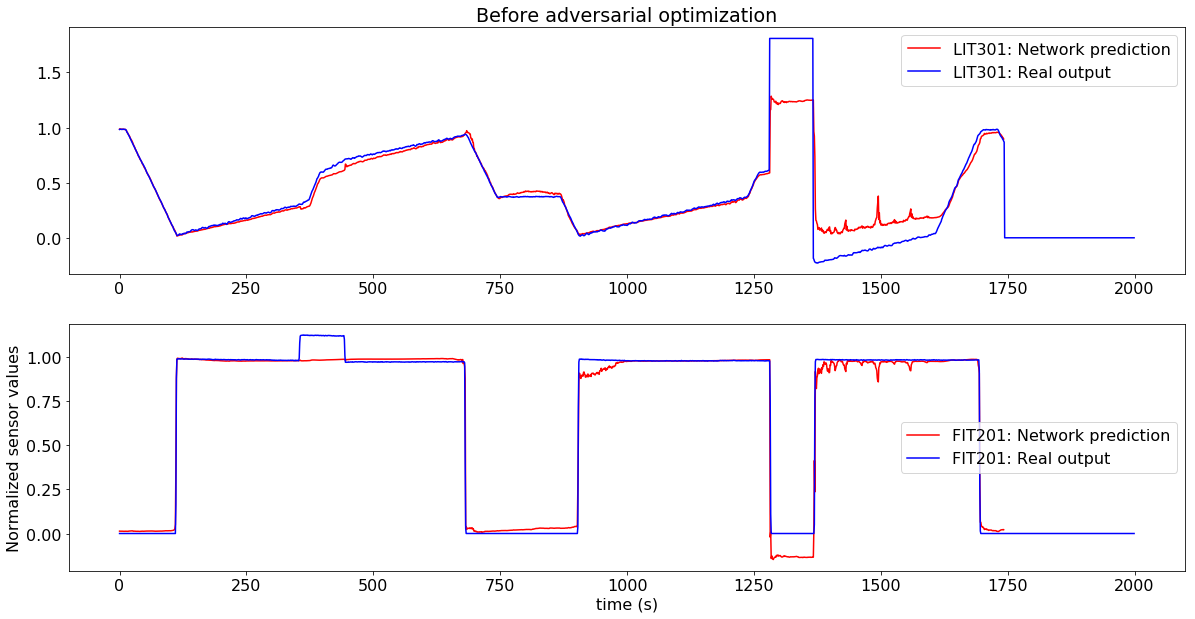}}
\caption{A 1D CNN model prediction for SWaT attack 7 with two fields. One can also see artefacts of another attack happening before attack 7.}
\label{fig:swat_attack_7_before_adv_2_fields}
\end{figure}
\begin{figure}[t!]
\centering{\includegraphics[clip, trim=0cm 0cm 0cm 0cm, scale=0.2]{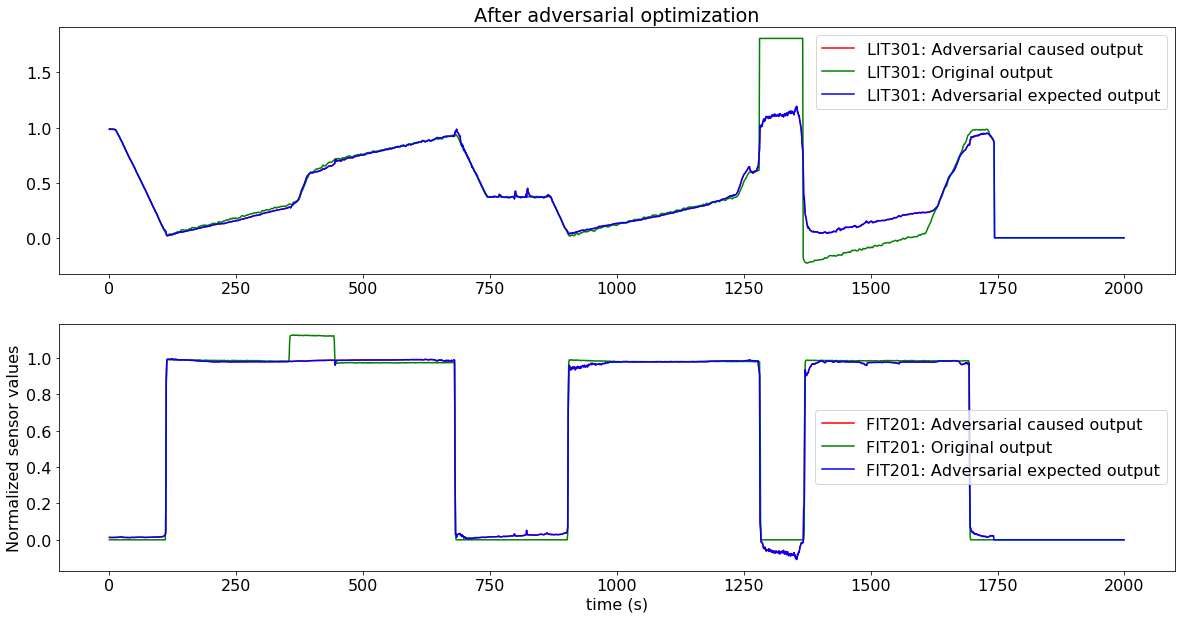}}
\caption{A 1D CNN model prediction for SWaT attack 7 after adversarial input optimization when using two fields. The adversarial input loses its desired physical impact}.
\label{fig:swat_attack_7_after_adv_2_fields}
\end{figure}

After the adversarial optimization we were able to produce input that retains the physical characteristics of the attack (maintaining the spoofed high level for the attack period) and was predicted by the model very closely, thus going undetected (see Figure~\ref{fig:swat_attack_7_after_adv}.
However, when we added a single additional feature to the model, the adversarial optimization undid the attacker's desired physical effect - the adversarial input conforms to the original model's prediction and does not cause underflow (see Figures~\ref{fig:swat_attack_7_before_adv_2_fields} and \ref{fig:swat_attack_7_after_adv_2_fields}).
The same behavior was observed with AE-based models.
In addition, we observed that using noise to corrupt the input, as described in Section \ref{sec:autoencoders} increases the robustness of the model to the adversarial evasion abilities even further, because the random noise applied to the adversarial examples is different between the adversarial training and testing time.

\begin{figure}[t!]
\centering{\includegraphics[clip, trim=0cm 0cm 0cm 0cm, scale=0.2]{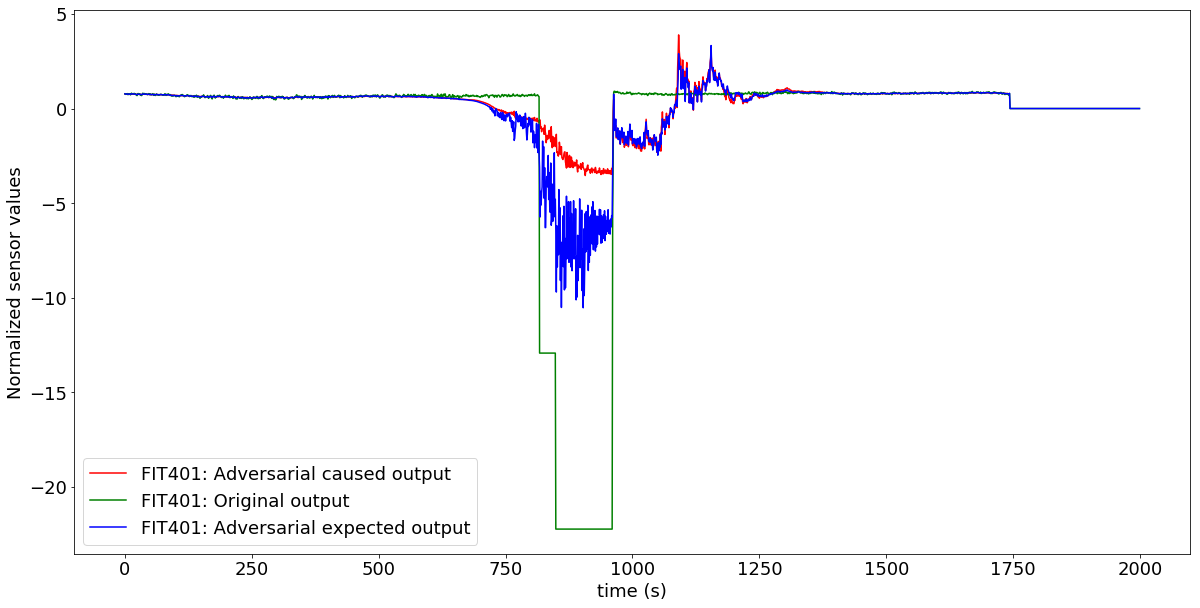}}
\caption{A 1D CNN model prediction for SWaT attack 10 after adversarial input optimization. The original attack shown in green spoofs the value of the flow reported by the sensor FIT401. The adversarial input aimed to cause the model to predict the values shown in red, but it failed to do so and reached the significantly different values shown in blue, while deviating from the intended physical impact significantly.}
\label{fig:swat_attack_10_after_adv}
\end{figure}

Generating adversarial inputs for other attacks demonstrated the same or even more robust behavior: sometimes it was not possible to create adversarial input that preserves the intended physical effect for a one-feature model (illustrated in Figure~\ref{fig:swat_attack_10_after_adv}).
In addition, if we constrain the level of allowed noise (e.g., to 0.05), the generation of adversarial inputs fails completely in all of our experiments. 

This adversarial robustness of physics-based models stands out against the background of successes of adversarial learning in other domains (e.g., image processing).
The most likely reasons for such robust behavior are the various constraints that the laws of physics, signal continuity, and the underlying PLC logic impose on the data and on the models.
In many areas, such as autonomous driving, the ability to trust the decision of neural networks is strongly influenced by the threat that the decisions were adversarially manipulated.
The results of our experiments suggest that the proposed anomaly detection method is resilient to adversarial evasion attacks that represent a major part of adversarial threats.

Despite the promising results, the data-only approach has limitations:
\begin{itemize}
    \item the tests were conducted only on the attacks present in the dataset; other attacks for which adversarial input could be created might exist, 
    \item the adversarial input generation relies on the model to represent the real system accurately, however as the attack conditions were never present in the training set, the actual system response might be different from the modeled one.
\end{itemize}
These limitations can be addressed by performing testing on a real system, which is a task for future research.

\section{Conclusions}\label{sec:conclusions}
In this paper, we studied the effectiveness of 1D CNN and AE-based anomaly and cyber attack detection mechanisms, answering our research questions as follows.
\begin{itemize}
    \item Based on our experiments, we conclude that both 1D CNNs and AEs achieve or exceed the state-of-the-art performance on the three public datasets, while maintaining generality, simplicity, and a small footprint. 
    It is not clear whether one of these architectures is always preferable over another, and we plan to extend our research with more datasets to investigate this further. 
    In the meantime, we recommend an ensemble consisting of both models when possible. 
    If a single model must be chosen, AEs will likely work out of the box in most cases, while 1D CNNs will require a round of hyperparameter tuning to eliminate false positives.
    \item We discovered that given the proper data preparation and feature selection, PCA-Reconstruction and windowed-PCA can provide a simple and efficient detector in many practical cases.
    Our recommendation is to first try PCA as a baseline detector before applying the neural network-based ones.
    \item The attack detection method we use allows us to pinpoint the attack location. 
    However, as spoofing attacks can  trigger a number of changes in the related features, all of the features will be considered attacked; in such cases, the attack area will be pinpointed still providing significant value to the operator. 
    Algorithmic means of distinguishing between PLC input and output require the integration of real-time network data, and this is a topic for future research.
    \item We found frequency domain analysis helpful in anomaly and attack detection. 
    Its applicability is subject to a number of practical limitations; if they are met, frequency domain analysis can provide strong results.
    \item The proposed detection method was found to be resilient to adversarial evasion attacks.
    This finding is a promising one as it allows the operators to trust the decisions of the method.
    There is a need to confirm these results with real testbeds. 
    An additional future research direction is a study of adversarial poisoning attacks on the proposed method.
\end{itemize}

\section{Acknowledgements}
The authors thank iTrust Centre for Research in Cyber Security, Singapore University of Technology and Design for creating and providing the SWaT and WADI datasets, Dr. Elchanan Zwecher for his valuable insights, and Rafael Defense Systems for supporting this work.

\bibliographystyle{IEEEtran}
\bibliography{refs} 

\vskip -2\baselineskip plus -1fil
\begin{IEEEbiography}[{\includegraphics[width=1in,height=1.25in,clip,keepaspectratio]{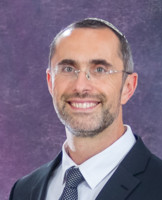}}]{Moshe Kravchik}
received his M.Sc. (2007) in computer science from the Open University of Israel.
He is currently pursuing a Ph.D. at
Ben-Gurion University of  the  Negev and teaches cyber security at Lev Academic Centre in Jerusalem. His research interests
include the topics of software and systems security, trusted execution environments, anomaly
detection, and the security of industrial control systems.
\end{IEEEbiography}
\vskip -2\baselineskip plus -1fil
\begin{IEEEbiography}[{\includegraphics[width=1in,height=1.25in,clip,keepaspectratio]{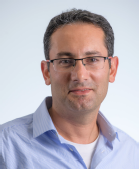}}]{Asaf Shabtai}
Asaf Shabtai is a professor in the Department
of Software and Information Systems Engineering
at Ben-Gurion University of the Negev. His main
areas of interest are computer and network security,
machine learning, security of IoT and smart mobile
devices, social network analysis, and security of
avionic and operational technology (OT) systems.
Asaf holds a B.Sc. in mathematics and computer
Sciences; B.Sc. in information systems engineering;
M.Sc. in information systems engineering and a
Ph.D. in information systems engineering all from
Ben-Gurion University.
\end{IEEEbiography}

\end{document}